# The Compositional Structure of the Asteroid Belt

F. E. DeMeo
*Massachusetts Institute of Technology*

C. M. O'D. Alexander
*Carnegie Institution of Washington*

K. J. Walsh and C. R. Chapman
*Southwest Research Institute*

R. P. Binzel
*Massachusetts Institute of Technology*

**Abstract:**
The past decade has brought major improvements in large-scale asteroid discovery and characterization with over half a million known asteroids and over 100,000 with some measurement of physical characterization. This explosion of data has allowed us to create a new global picture of the Main Asteroid Belt. Put in context with meteorite measurements and dynamical models, a new and more complete picture of Solar System evolution has emerged. The question has changed from "What was the original compositional gradient of the Asteroid Belt?" to "What was the original compositional gradient of small bodies across the entire Solar System?" No longer is the leading theory that two belts of planetesimals are primordial, but instead those belts were formed and sculpted through evolutionary processes after Solar System formation. This article reviews the advancements on the fronts of asteroid compositional characterization, meteorite measurements, and dynamical theories in the context of the heliocentric distribution of asteroid compositions seen in the Main Belt today. This chapter also reviews the major outstanding questions relating to asteroid compositions and distributions and summarizes the progress and current state of understanding of these questions to form the big picture of the formation and evolution of asteroids in the Main Belt. Finally, we briefly review the relevance of asteroids and their compositions in their greater context within our Solar System and beyond.

## 1. Introduction:

By the early 1990s in the era of *Asteroids II*, roughly 10,000 asteroids had been discovered, only a fraction of the total number of asteroids that are now known to exist. At that time, asteroids were discovered by visually inspecting photographic plates for light trails. In the 1990s, many of the major automated discovery surveys came online. By the year 2000 around the time of *Asteroids III*, ~20,000 asteroids were known (see Figure 1). Today (in 2015, *Asteroids IV*) there are roughly 700,000 asteroids with known orbits, revealing much new information about the Asteroid Belt's dynamic past. About 100,000 asteroids have measurements that tell us about their surface compositions (*Ivezic et al.*, 2001, *Szabo et al.*, 2004, *Nesvorny et al.*, 2005, *Carvano et al.,* 2010), providing a broader view of the Asteroid Belt than ever before. With the explosion of asteroid discoveries over the past decade (Fig. 1) and the conclusion of two of the largest asteroid physical measurement surveys, the Sloan Digital Sky Survey (SDSS; *Ivezic et al.,* 2001) and the Wide-field Infrared Survey Explorer (WISE; *Mainzer et al.,* 2011), it is timely to reflect upon recent advances, leaving us well poised for the next generation of major surveys including Gaia (*Mignard et al.*, 2007) and the Large Synoptic Survey Telescope (LSST; *Ivezic et al.,* 2007, Jones et al., 2009).

This Chapter reviews our current understanding of the compositions of asteroids in the Main Belt from asteroidal, meteoritic, and dynamical perspectives. Our view of the Asteroid Belt has changed dramatically since the "big picture" chapters from *Asteroids I* and *II* (*Chapman* 1979, *Gradie et al.,* 1989, *Bell* 1989). In the context of this chapter, the use of the term "compositional trends" is more akin to taxonomic rather than mineralogical trends and the "compositions" discussed here refer more to broad trends of observational data than any derived mineralogical information. In Section 2, we review the current tools used to compositionally (taxonomically) characterize and classify asteroids and meteorites, and the dynamical tools that help us interpret the current orbital distribution of the asteroids. We also review the current distribution of asteroid classes in the main Asteroid Belt. Section 3 focuses on how the current observational constraints strengthen or weaken leading dynamical theories. In Section 4, we summarize the meteoritic evidence for asteroid compositions, as well as when and how asteroids formed. In Section 5, we compile the major compositional questions in asteroid science and review the progress made toward answering many of them to provide a broad view of asteroid compositions and their locations in the Belt. Section 6 looks at asteroids in their greater context for Earth, the Solar System, and beyond. The Appendix lists the major outstanding questions relevant to asteroid compositions and distributions. The Appendix also briefly summarizes the current state of knowledge and suggests future work to solve these problems. Many of these questions are noted specifically in this chapter and are addressed, and most are covered in more detail in other chapters.

Figure 1: Asteroid Discoveries

## 2. The Observational Perspective

### 2.1 Asteroid Composition Tools

The surface properties (grain size, mineralogy, degree of space weathering, etc.) of an asteroid can be inferred through spectral and photometric measurements at wavelengths from the ultraviolet (UV) to the infrared. Thermal emission at longer wavelengths is used to calculate surface albedos that are related to surface compositions (see chapter by *Mainzer et al.*). Absorption features in reflectance spectra from UV to mid-infrared wavelengths and emission features in the mid-infrared range can be used to identify minerals and other compounds on the surface of an asteroid or meteorite (see chapter by *Reddy et al.*). For example, olivine and pyroxene have readily identifiable absorption features located at one and, for pyroxene, two microns.

While mineralogical analysis is most appropriate for detailed studies of bodies with distinct features, relatively few asteroids have the high quality spectra that are required for this. On the other hand, taxonomic classifications can be made using lower quality or lower resolution spectra, providing a rapid characterization of asteroid spectra and a common language for comparing them. Hence, taxonomic information is available for orders of magnitude more than the number of asteroids for which detailed spectra have been measured. At the time of *Asteroids III,* the majority of reflectance spectra were taken at visible wavelengths to one micron, and were classified according to the Bus or Tholen taxonomies (*Bus & Binzel* 2002b, *Tholen & Barucci, 1989*). The mafic silicate-rich asteroids with available near-infrared spectra at the time were classified mineralogically by the Gaffey system (*Gaffey et al.*, 1993). By the early 2000s, near-infrared spectrometers became available, such as SpeX on the NASA IRTF (*Rayner et al.,* 2003). An extension of the Bus taxonomy, the so-called Bus-DeMeo taxonomy, was created to classify both the visible and near-infrared data in such a way as to be as consistent with the Bus taxonomy as possible (*DeMeo et al.,* 2009). A comparison of each of these major taxonomies is presented in Table 1.

Table 1 Near Here

Asteroid spectra are traditionally divided into three major complexes and each of the complexes is divided into individual classes (also called "types"). The S-complex, originally named for its expected silicaceous composition (*Chapman et al., 1975*), is characterized by spectra with moderate silicate absorption features at 1 and 2 microns. The C-complex, historically named in connection with carbonaceous chondrite meteorites, have low albedo surfaces with spectra that have flat or low slopes and are subtly featured to featureless. Subtle features have absorptions of only a few percent, one of the most notable being the 0.7-micron feature indicating the presence of phyllosilicates likely due to aqueous alteration (*Vilas & Gaffey 1989*). The X-complex is characterized by moderately sloped and subtly-featured or featureless spectra. It has long been known that the X-complex is compositionally degenerate because it comprises the darkest and brightest surfaces of all asteroids, with albedos as low as a few percent to as high as 50 %. The Tholen taxonomy distinguished the X-complex by albedo, breaking it up into the E, M, and P classes that ranged from high to low albedo.

Additionally, there are spectral classes that do not fit within these three main complexes. These spectra typically have more extreme or distinct characteristics and are much less common than the major complexes. These include the spectrally featureless and very red-sloped D-types, the spectrally olivine-dominated A-types with the broad, deep one-micron feature, the spectrally pyroxene-dominated V-types with a deep, narrow one-micron feature and a deep, broad two-micron feature. Additionally, there are the K,

L, T, O, Q, and R-types. A visual summary of the visible and near-infrared spectral classes is provided in Figure 2. A description of the spectral features can be found in *Bus & Binzel 2002b* and *DeMeo et al., 2009*.

Figure 2: Asteroid Taxonomy

The accuracy of an asteroid's classification depends on the quality of the data. Also, repeated observations of a single object may show spectral (hence perhaps taxonomic) differences that may be intrinsic (due to rare mineralogical or grain size variations on different portions of the body), may reflect real but uncorrected variations due to photometric viewing geometry, or may reflect instrumental or observational artifacts. We emphasize the advice from the chapter by *Bus et al.,* in *Asteroids III*, "We should not feel compelled to decide which label is 'correct,' but should rather accept these distinct labels as a consequence of our growing knowledge about that object." For example, an object classified by visible wavelength measurements alone may 'change its type' when the wavelength coverage is extended to the near-infrared, or when an albedo measurement becomes available. An improved answer as more information becomes available is simply the scientific method at work.

One of the biggest challenges remains how to determine surface compositions from asteroid spectra and spectral types. Most materials on an asteroid's surface do not produce distinct, identifiable features in a reflectance spectrum. Furthermore, an asteroid spectrum is not only a product of composition, but is also affected by factors such as grain size, temperature, viewing geometry, and space weathering processes (see chapter by *Reddy et al.*). The compositional interpretation of many asteroid classes has broadened over time. While the S-complex has been linked to ordinary chondrite meteorites and the C-complex to carbonaceous chondrites, there is strong evidence that these spectral classes are not compositionally uniform. There is a range of different types of ordinary and carbonaceous chondrites, and there are additional meteorite classes that also may comprise subsets of the S- and C-complexes. This must be accounted for when interpreting spectra.

In addition to the visible and near-infrared, asteroid spectra are now also measured beyond 2.5 microns and into the mid-infrared. Asteroid spectra at these wavelengths are categorized by the presence of and shape of a 3-micron absorption feature indicating hydrated minerals or phyllosilicates (*Takir & Emery 2012, Rivkin et al., this volume*). There are four distinct spectral groups of 3-micron features. Beyond the 3-micron feature, the mid-infrared range (7-25 microns) holds a wealth of information (*Emery et al., 2006*) because many minerals have diagnostic features at these wavelengths. There are multiple deterrents to analyzing these longer wavelengths: instrument design sensitivity requirements are challenged by measurements taken at ambient temperatures; observations are constrained by the lower solar flux; and available laboratory experimentation to interpret the spectra is currently limited in number (e.g. *Vernazza et al., 2010, 2012*).

## 2.2 The Historical Perspective

Initial measurements from the 1940s through the 1970s found that the surface brightness and colors of asteroids trend from medium albedo and moderate spectral slopes (S-complex) for bodies in the inner part of the Main Asteroid Belt to lower albedo and neutral spectral slopes (C-complex) toward the outer part of the Main Belt (*Fisher* 1941, *Kitamura* 1959, *Wood & Kuiper* 1963, *Chapman et al.,* 1971, *Chapman et al.,* 1975). The diversity was a great surprise and the trend with distance motivated further observations.

The hundreds of observations available with even greater wavelength coverage in the 1980s then revealed distinct peaks in relative abundance for each of the major classes of asteroids at certain distances throughout the Belt (*Zellner et al.,* 1985, *Gradie & Tedesco* 1982, *Gradie et al.,* 1989). Because the asteroids were grouped systematically with heliocentric distance, it was concluded that the asteroids formed close to their current locations. In the context of a relatively static Solar System, where the asteroids were assumed to have formed nearly in place, the variation in the compositions of these asteroids was interpreted to represent the original thermal gradient across the Main Belt from the time of planetesimal formation (see question C1 of Appendix A). While this fundamental trend among the largest asteroids has remained robust, both our interpretation of surface properties and the dynamical models explaining this structure have evolved.

Two major surveys from the late 1990s to early 2000s (*Bus & Binzel 2002, Lazzaro et al., 2004*) provided spectroscopic data for around 2,000 asteroids. While much of the ensuing analysis focused on asteroid families, two works investigated taxonomic distributions with orbital properties. *Bus & Binzel* (2002) note a more detailed distribution than *Gradie & Tedesco*, finding double peaks for the S- and C-complex with distance from the Sun. *Mothe-Diniz* et al., (2003) found that the distribution of C-complex objects was different for higher versus lower inclinations. These authors were also the first to show the presence of S-complex asteroids at distances as far as 3 AU in their dataset of asteroids larger than 15 km and found a fairly even distribution of S-complex out to that distance.

**2.3 Recent Progress**

The next major survey released was the Sloan Digital Sky Survey. Although it was designed for extragalactic studies, it made crude spectral observations of over a hundred thousand moving objects in the field. These data were released in several Moving Object Catalogs (*Ivezic et al., 2001, Parker et al., 2008*). Combined with the more recent NEOWISE survey that produced over a hundred thousand diameters and albedo estimates (*Mainzer et al., 2011, Masiero et al., 2011*) there now exists a treasure trove of physical measurements to analyze. Many of the advances in understanding the compositions of asteroids in the Main Belt since *Asteroids III* have stemmed from these datasets, as well as targeted spectroscopic surveys or targeted follow up studies of objects in these two large datasets. Here we describe taxonomic trends beginning in the Hungaria region and extending out to the Jupiter Trojans. Figure 3 shows the distribution of the asteroid class in each heliocentric distance region and multiple size regimes. We use the terms "Inner," "Middle," and "Outer" Main Belt to refer to the three regions of the Main Belt carved by the $\upsilon6$ secular, 3:1 and 5:2 mean-motion resonances (the $\upsilon6$ starts near 2 AU and the 3:1 and 5:2 are located at 2.5 and 2.82 AU, respectively).

The Hungaria region (1.8-2.0 AU, i~20 degrees) is dominated by high albedo (>0.3, *Tholen & Barucci, 1989*) E-type asteroids that have moderate spectral slopes and often display a 0.49-micron absorption feature (*Gradie & Tedesco 1982, Bus & Binzel 2002b*). These E-types are considered members of the Hungaria asteroid family (*Gaffey* *et al.*, 1992) with (434) Hungaria being the brightest member. In the Hungaria region, however, there are a variety of compositional types, including S- and C-complex objects (*Carvano et al., 2001*).

In the Inner Main Belt (2.0-2.5 AU), the dominant players are (4) Vesta (V-type) and a number large S-complex asteroids. C-complex are rare in the Inner Belt at large sizes (D>100 km) where they comprise only 6% of the total mass, but they make up a quarter of the mass at medium sizes (20 km<D<100 km), and are almost equal to the S-complex by mass at the smallest sizes (5 km<D<20 km). At the same time, the fractions of medium-sloped spectral types (M and P) decrease at smaller sizes. Newly discovered in the Inner Main Belt are D-type asteroids defined by their very red spectral slopes, which had only previously been seen at larger distances aside from a few NEOs that have dynamical origins in the Outer Belt and beyond (*Carvano et al., 2010, DeMeo & Carry 2014, DeMeo & Binzel 2008*). These objects are discussed later in Section 5.2.

An interesting effect of viewing the Inner Main Belt by mass (previous analyses have viewed such statistics by number, not by mass) is the relative insignificance of the Vesta family, the products of a large collision with Vesta, in the Inner Belt (*Binzel & Xu* 1993). Indeed, only a handful of all Vesta family members, called Vestoids, are larger than 5 km, so their mass contribution even among 5-20 km diameter bodies is miniscule (1% of that size range and region, DeMeo & Carry 2014). Vestoids are significant contributors to the Inner Belt in terms of the total number observed (*Parker et al.*, 2008, *Masiero et al.*, 2013), but it is their high albedo, close distance, and spectral distinctiveness that have biased their discovery and classification.

In the Middle Main Belt (2.5-2.82 AU), Ceres (C-type in the Bus-DeMeo taxonomy) and Pallas (B-type) are the largest objects and they comprise roughly 31% and 7%, respectively, of the entire Main Belt by mass. The broad taxonomic makeup of the Inner and Middle Belt at the smallest sizes is essentially identical.

In the Outer Main Belt (2.82-3.3 AU), the C-complex dominates by mass with Hygeia being the largest and most massive member. Despite the fact that the relative fraction of S-complex asteroids is small in the Outer Main Belt, their total mass is still quite significant given that the mass in the Outer Belt is 2-10 times greater than in the Inner Belt at each size range.

A- and V-types, respectively olivine-dominated and basaltic asteroids, are present in small numbers throughout the Main Belt, aside from those associated with Vesta (*Lazzaro et al., 2000, Moskovitz et al., 2008, Sanchez et al., 2014*). Their discovery in the Middle and Outer Belts was surprising since differentiated bodies or fragments of them were not expected in the context of the classical understanding of asteroid differentiation. Significant advances have been made since Asteroids III in understanding the complexity of both the asteroid differentiation process and of the mechanisms that displace material through out the Solar System (see chapters by *Scheinberg et al., Scott et al., Morbidelli et al.*).

Families play a very important role in the architecture and composition of the Main Belt. The chapter by *Masiero et al.,* covers this topic in detail. It will be valuable for future work to incorporate the size-frequency distribution of asteroid families into studies of the radial distance distribution of asteroids in the Main Belt to fully interpret the results, particularly at small asteroid sizes.

The taxonomic makeup of the largest Hildas (~4 AU) and Trojans (5.2 AU) remain predominantly P-type and D-type, respectively. The trends at smaller sizes are discussed in Section 5.2. The chapter by Emery et al., is dedicated to the Trojan population, covering the compositional and physical characteristics as well as the dynamical history.

**Figure 3 Here**

## 3. The Dynamical Perspective

### 3.1 Dynamical Tools

Determination of an asteroid's orbit immediately tells you where it is currently spending its time in the Solar System, and the combination of the distributions of known orbits and physical properties provide powerful clues to the evolution of the Solar System (Appendix Questions A2 and A3). However, there are many more asteroids with known orbits than there are with physical characterization, and at times we are left to gain context based on their orbits alone. The dominant perturber in the Solar System is Jupiter and its effects are made clear by the large depleted Kirkwood gaps in the asteroid population owing to its mean motion resonances. Given the dominance of Jupiter's perturbations for asteroid orbits, one can take advantage of its similarity to the restricted three-body problem to generate some quasi-conserved quantities. The most common measure is the Tisserand parameter $T_J$, which is calculated with respect to Jupiter and can help to distinguish between different classes of small body orbits. This is primarily used to separate Jupiter Family Comets ($2 < T_J < 3$) from nearly isotropic comets ($T_J < 2$), but is also commonly used to try to uncover dormant comets in the NEO population (*Levison et al.*, 1994, *Bottke et al.*, 2002b, *DeMeo & Binzel* 2008) and separate Main Belt Comets from the ordinary comet population (*Jewitt et al.,* this volume).

Orbits are typically calculated and reported for what the asteroid's Keplerian orbit would be at a specific epoch in the presence only of the Sun. This calculated "osculating orbit" does not incorporate any information about short or long-term oscillations of the orbit owing to the perturbations of the giant planets. The "proper elements" of an asteroid represent quasi-integrals of motion, meaning that they are nearly constant in time, and can be calculated or estimated using various numerical and analytical tools (*Knezevic et al., 2002*). The difference between an asteroid's osculating and proper orbital elements can be substantial – tens of percent in *a, e* and i. The proper elements of an asteroid are representative of its long-term orbit and are essential in studies of dynamically related clusters or families of asteroids (see *Nesvorný et al., in this volume*, *Bendjoya et al., 2002, Zappala, 2002*).

An important dynamical process affecting the entire Main Asteroid Belt is the Yarkovsky effect (see *Bottke et al., 2002a* and *Bottke et al., 2006*). This is a size-dependent drift in a body's semimajor axis caused by the re-emission of absorbed solar radiation. It is the main driver that pushes Main Belt asteroids into resonances and can then become NEOs or be driven to extreme orbits that could have them impact a planet or the Sun, or be ejected from the Solar System. Smaller objects move more rapidly, where drift rates scale roughly as $1/D$, which means that where we find smaller objects now may simply be a waypoint on a drift across the Asteroid Belt (where a 1 km body might drift roughly $10^{-4}$ AU in 1 Myr). A similar point of confusion for small objects is that the collisional lifetimes are much shorter than the age of the Solar System at sizes smaller than 10-30 km (*Bottke et al., in this volume*). Between the effects of the Yarkovsky drift and collisional evolution, we can really trust the orbits of only the 500 or so largest bodies (D> 50 km) as tracers of the early structures of the Asteroid Belt.

As the Yarkovsky effect pushes asteroids around, some will inevitably reach an orbital resonance (a mean motion resonance with Jupiter or other powerful secular resonances; see *Nesvorny et al.*, 2002). While the effects of this may vary, a typical response is an increase in the asteroid's orbital eccentricity. As the eccentricity increases to large values it can cause the body to cross the orbit of Jupiter, which can easily result in its ejection from the Solar System. If the asteroid first has an interaction with the terrestrial planets it is possible that its orbit can be altered in a way to pull it out of the Main Asteroid Belt and reside almost entirely within the inner Solar System. In near-Earth space, the orbital dynamics are controlled by the chaotic interactions with the terrestrial planets and thus the population is transient, with only a 10 Myr lifetime (*Gladman et al., 2000*). Numerical models of the evolution from the Main Belt to near-Earth orbits have largely recovered the flux of bodies moving through different resonant passageways and can explain the size of the NEO population (*Bottke et al., 2000, Bottke et al., 2002b*). Furthermore, the NEO model of *Bottke et al., 2002b* provides a dynamical tool to make statistical links between NEAs and the most likely resonant pathway they traveled from the Main Asteroid Belt. Some NEOs can be probabilistically linked to specific regions of the Main Belt, allowing for links to be made between bodies and their regions or possible parent asteroid families (e.g., *Campins et al., 2010*).

Even the orbits of the largest bodies have likely been altered substantially early in Solar System history. As models of Solar System evolution have matured, the effects of the possible "late" (after ~400 Myr or so) movements of the giant planets have been studied in the most detail (see *Morbidelli et al., in this volume*). Specifically, some traces of giant planet migration are still found in the Main Asteroid Belt by way of depletions near primordial mean motion resonances with Jupiter (*Minton and Malhotra 2009*), though some types of migrations can be ruled out by clear patterns of depletion and orbital changes that would still be noticeable in today's Asteroid Belt (*Morbidelli et al., 2010*). As the possible late migration of the giant planets has been studied in more detail, it is clear that many small bodies could be affected. The violent instability of the giant planets has been found to provide ideal dyanamical pathways to capture the Trojan asteroids at Jupiter (*Morbidelli et al.,* 2005, *Nesvorny et al.,* 2013), capture the irregular satellites of the giant planets (*Nesvorny et al.,* 2014), implant D- and P-type asteroids from the primordial Kuiper Belt into the Main Asteroid Belt (*Levison et al.,* 2009) and sculpt the Kuiper Belt (*Levison et al.,* 2008, *Batygin et al.,* 2012). An example of the effects of planetary migration would have on small bodies and the asteroid belt is shown in Fig. 4.

The distribution of asteroid orbits, their physical properties and the total mass depletion in the Asteroid Belt region are all used as constraints on early Solar System evolution – as discussed below (and in more detail in the chapter by *Morbidelli et al.*).

### 3.2 Dynamical Overview of the Main Belt

The Asteroid Belt today is estimated to contain approximately ~5×10$^{-4}$ Earth masses, which is approximately three times the mass of its most massive asteroid (1) Ceres (*Krasinsky et al.*, 2002). This is in contrast to the nearly one Earth mass of material that would be expected to inhabit this region given a smooth distribution of the solid material found in the planets (*Weidenschilling* 1977). Similarly, classical models of planetesimal formation suggest that the current mass in the Asteroid Belt today would not be enough to have grown the largest asteroids and thus there has been a depletion of mass since the Asteroid Belt formed (see *Weidenschilling* & Jackson, 1993). As discussed in detail in the chapter by *Morbidelli et al.*, (and below), there are multiple proposed methods to remove so much mass from the Asteroid Belt, and there are also new models for planetesimal formation that may allow the formation of large asteroids directly from small pebbles in the solar nebula (*Johanson et al.*, in this volume).

The orbital distribution of the Asteroid Belt finds that most of the dynamically stable phase space of *a,e,i* is filled, with orbital eccentricities ranging from 0 to 0.35 and inclinations from 0 to about 30 degrees. Meanwhile, every theory for planetesimal formation relies on the damping effects of the gaseous solar nebula to reduce relative velocities and increase accretion rates (see *Johansen et al.,* chapter in this book).. This implies that, at least immediately following their formation, planetesimals would have been on circular and co-planar orbits and were dynamically excited into their current orbits at some later time.

The loss of mass, the observed dynamical excitement and taxonomic mixing are presumed to be closely linked to the dynamical evolution of the Solar System – including the formation and migration of the terrestrial and giant planets. As described in more detail in the chapter by *Morbidelli et al.*, models typically attack all three constraints with one mechanism. Studies of sweeping secular resonances due to the dissipation of the solar nebula suffer dual problems of needing long (20 Myr) timescales of gas dissipation to deplete enough bodies, and then typically fail to reproduce the inclination distributions (see *O'Brien et al.,* 2007). The inclination distribution is also a problem for models that invoke stranded planetary embryos that excite and deplete the primordial Asteroid Belt (see *O'Brien et al.,* 2006). Here there is depletion of low inclination bodies owing to their low-velocity encounters with planetary embryos, and again a mis-match with the observed distributions. Collisional processes are also important constraints. While even very massive Asteroid Belts (e.g. with 1000 times more mass than today) would have collisionally ground away most mass and reached a total mass similar to that found today (Chapman & Davis 1975), the remnants and scars of such dramatic and long-term collisional evolution would likely be more visible in our studies today (see also the chapter by *Bottke et al.).*

Partly due to models inability to simultaneously match constraints with the terrestrial planets (the mass of Mars) and with the Asteroid Belt (orbital distributions and water delivery), a recently proposed model, the Grand Tack Model, invokes a scattering implantation of nearly the entire Asteroid Belt population from different parent populations (*Walsh et al.*, 2011). This dramatic migration of the giant planets causes widespread depletion and then mixing of remnant populations in the dynamically stable Asteroid Belt. When Jupiter is migrating inwards it completely depletes all objects native to the current Asteroid Belt. During its outward migration it scatters some remnants of this population back into the Asteroid Belt, and then during the outermost stretches of its migration it also scatters in bodies from more primitive reservoirs between and beyond the formation region of the Giant Planets. This mechanism is distinct from others as it implies separate parent populations for some of the major different compositional classes found in the Asteroid Belt, and also because it results in a low-mass Asteroid Belt from very early on in Solar System history. While it provides first-order matches to these three Asteroid Belt constraints (mass depletion, orbital and taxonomic distributions), they were not a prediction of the model – rather they were a necessary constraint for the model to be viable. Going forward, each of these can be investigated more closely and hopefully limit or rule out some of the free parameters in the current Grand Tack scenario (growth and migration parameters of the giant planets).

Already, the Grand Tack relies on some assistance from what is thought to follow in Solar System history. The eccentricity distribution in the Grand Tack model is elevated compared to what is observed today (*Walsh et al.*, 2012). However, the motion of the giant planets during the "Nice Model", expected to happen roughly 400 Myr later, will alter the orbits of the Asteroid Belt while having minimal affect on taxonomic distributions or total mass (eccentricities are the primary orbital element altered, and total mass is only depleted by a factor of 2-3; *Morbidelli et al.*, 2010 --- though the Nice Model is credited with implanting D- and P- types bodies in the Main Belt). In fact, *Minton and Malhotra* 2011 find that a very excited Asteroid Belt (elevated eccentricities) is a good fit when considering very simple models of giant planet migration that could come later. However, the models regarding later giant planet migration is a field of active study, and so each new iteration may require a reinvestigation of this aspect of the Grand Tack's Asteroid Belt fits.

If the Grand Tack relies on later events, it begs the question of how much we can use different mechanisms to explain different constraints – and are there ways to mesh Asteroid Belt constraints with other models of planet formation? Already planet formation models are increasingly using the Asteroid Belt as a constraint for their model outcomes, both for delivering water-rich asteroid material to the growing planets (*Morbidelli et al.*, 2000), and also for using their orbits to rule out other modes of planet migration (*Minton and Malhotra* 2009, *Walsh and Morbidelli* 2011).

Meanwhile recent advances in planetesimal formation (see chapter by Johansen et al.) imply the possibility of a different initial distribution of mass in the early Solar System than previously considered. It is possible that planetesimal formation relies on clumps of "pebbles" collapsing, which could lead to a small number of planetesimals amidst a huge number of remnant pebbles – where only a few of the formed planetesimals are large enough to rapidly accrete the remaining pebbles (see *Ormel and Klahr* 2010). One could envision scenarios where the Asteroid Belt never had much mass, and thus dynamical and taxonomic stirring would be constraints independent of mass loss.

## 4. The Meteorite Perspective

### 4.1 Meteorite Composition tools

The parent asteroids of most meteorites would have formed at different times and/or places in an evolving solar nebula. This will have had profound effects on the initial compositions and subsequent histories of the asteroids that are reflected in the way meteorites are classified.

Radial thermal gradients in the disk will have dictated gross differences in compositions upon accretion, such as rock/ice ratios. However, it is evident from the meteorite record that more transient processes were also important in the thermal processing of dust and that radial transport brought together materials with different thermal histories. The thermal processing of dust has left its imprint on the major and trace element compositions that show clear variations associated with their volatility. Estimates of the relative volatilities of the elements are traditionally based on thermodynamic equilibrium calculations of their 50% condensation temperatures from a gas of solar composition at a total pressure of $10^{-4}$ bars (e.g., *Lodders, 2003*). There is also evidence in meteorites for the fractionation of elements in the nebula according to their chemical affinities (lithophile – rock-loving, siderophile – Fe-metal-loving, and chalcophile – sulfide-loving). Physical processes that separated solid/melt from gas and silicates from metal seem the most likely causes for these variations.

After accretion, asteroids were subject to internal heating, largely due to the decay of the short-lived radionuclide $^{26}Al$ ($t_{1/2}$≈720,000 years) that was inherited from the protosolar molecular cloud. Thus, asteroids that formed early will tend to have been more internally heated than those that formed later, although other parameters will also have influenced internal temperatures. The least heated meteorites (<150 °C) normally exhibit varying degrees of aqueous alteration. Thermally metamorphosed meteorites have been more severely heated (up to 800-900°C), which will have dehydrated them if they were not dry already, and they exhibit varying degrees of recrystallization and chemical re-equilibration between minerals. Finally, there are meteorites that came from bodies that underwent melting and, in many cases, differentiation that produced metal-sulfide cores and silicate mantles.

The classification of meteorites has recently been reviewed in considerable depth (*Krot et al., 2014*), and so is only briefly summarized here. The most basic separation of meteorites is into unmelted (chondrites) and melted ones (non-chondrites). The chondrites are generally assumed to come from parent bodies that were smaller and/or formed later that those of the non-chondrites, although it is possible that some chondrites are the unmelted crusts of differentiated bodies (*Weiss and Elkins-Tanton, 2013*). For further subdividing meteorites, the most useful classification tools reflect the meteorites' nebular (primary) and parent body (secondary) characteristics – both physical and chemical. It is important to note that the causes of the variations in primary and secondary features are not always well understood, but that makes them no less useful as classification tools. Superimposed on the primary and secondary characteristics there can be shock features and brecciation associated with later large impacts, as well as terrestrial weathering. A complete classification scheme must also account for these effects, but they will not be considered here.

For classification based on bulk chemistries, it is usually sufficient to use a few representative element ratios that reflect the major fractionations that influenced meteorite compositions. The CI chondrites enjoy a unique status among meteorites because their bulk compositions are identical within error to the rock-forming component of the solar photosphere. Since all objects are thought to have ultimately evolved from the solar composition, elemental compositions and ratios are usually normalized to (divided by) the CI composition.

Oxygen isotopes have proved to be a very useful classification tool because, in addition to the mass dependent fractionations produced by most physical and chemical processes, meteorites exhibit mass independent variations (changes in $^{16}O$ abundance relative $^{17}O$ and $^{18}O$). Recently, other isotope systems, most notably Cr, have begun to be used as additional classification tools.

A significant number of meteorites do not fit easily into the established groups. Since by convention groups must be composed of five or more members, these meteorites are classified as ungrouped. Each meteorite group is generally assumed to come from a single asteroidal parent body that had a uniform composition at the time of its formation. Nevertheless, it cannot be ruled out that members of a group come from a number of asteroids that formed at similar times and places (e.g., *Vernazza et al.*, 2014; Appendix Question B2). Assuming that a group comes from a single parent body and that ungrouped meteorites come from separate parent bodies, meteorites appear to be samples of 100-150 distinct parent bodies (*Burbine et al.*, 2002; Appendix Question B1).

### *Chondrites*

The chondrites are composed of three major components – refractory inclusions and chondrules, both of which formed at high temperatures, are embedded in a fine-grained matrix. Refractory inclusions are a diverse group of objects, broadly divided into Ca-Al-rich inclusions (CAIs) and amoeboid olivine aggregates (AOAs). The inclusions are the oldest known Solar System solids and, in some cases, formed by condensation from a hot gas, while others formed by melting and evaporation of pre-existing solids (*MacPherson*, 2014). Chondrules appear to have typically formed 1-3 Myr after refractory inclusions as molten silicate-metal droplets in brief (hours to days) heating events in very 'dust-rich' environments. Being so fine grained, the matrix is very susceptible to parent body modification, but seems to originally have been dominated by a mix of fine-grained (<5-10 μm) crystalline and amorphous silicates. Matrix also contains the bulk of the most volatile elements in chondrites, all of the organic matter and presolar circumstellar grains found in them, and probably originally contained the ices accreted by most chondrite groups.

Almost all chondrites belong to one of five classes that have been further subdivided into a number of groups (Table 2). Each group is defined by a narrow range of primary physical features (e.g., chondrule sizes, abundances, types, etc.), elemental compositions and O (and Cr) isotopic compositions.

After accretion, varying degrees of aqueous alteration and/or thermal metamorphism in the parent bodies led to the modification of the primary features of all chondrites. The extent of this modification is reflected in a chondrite's petrologic type - a chondrite that was unaffected by either aqueous alteration or thermal metamorphism is assigned a petrologic type of 3.0, while petrologic types 3-6 reflect increasing degrees of thermal metamorphism and types 3-1 reflect increasing extents of aqueous alteration. A meteorite's petrologic type is normally given after its chemical group (e.g., CI1, LL3.0 and H6).

Table 2 Here

***Non-Chondrites***

As with the chondrites, the non-chondritic meteorites are classified according to the properties of their components (mineralogy, grain size, etc.), as well as by their bulk chemical and isotopic compositions (Table 3). Except for those from the Moon and Mars, it is generally assumed that non-chondritic meteorites evolved from originally chondrite-like objects. They are divided into (i) primitive achondrites that have seen very intensive metamorphism, as well as relatively low degrees of partial melting and melt extraction, and (ii) differentiated meteorites that underwent extensive melting and differentiation. The differentiated meteorites are further divided on the basis of their Fe-metal contents into achondrites (metal-poor), stony-irons, and irons.

Primitive achondrites sometimes contain relict chondrules, which are chondrules that have survived metamorphism and partial melting of its host rock. However, they have not been conclusively linked to any known chondrite groups. Only one primitive achondrite, the ungrouped Tafassasset, has geochemical affinities to carbonaceous chondrites (*Gardner-Vandy et al.,* 2012). Similarly, only one achondrite, the ungrouped basaltic achondrite NWA 011and several paired specimens from the same fall, seems to have formed by melting of a carbonaceous chondrite-like body (*Warren*, 2011).

*Stony irons* – Mesosiderites are breccias composed of diverse silicate clasts that are intimately mixed with roughly equal amounts of Fe,Ni-metal and sulfide. Pallasites are composed of Fe,Ni-metal/sulfide that is intergrown with similar proportions of relatively coarse grained olivine and pyroxene crystals. There are three recognized sub-types that probably came from at least three different parent bodies. Traditionally, it has been assumed that pallasites are core-mantle boundary samples. However, recent studies indicate that the main-group pallasites experienced a range of cooling rates, which would not be expected if they come from the core-mantle boundary of a single body (*Yang et al.*, 2010b).

*Irons* – Iron meteorites are divided into 14 chemical groups, although ~15% of irons remain ungrouped. The chemical groups fall into one of two categories, magmatic and non-magmatic. The magmatic irons exhibit intragroup chemical variations that are consistent with fractionation between solid and liquid metal that presumably occurred during crystallization of planetesimal cores. The chemical variations within the non-magmatic iron groups are more difficult to understand and they tend to contain abundant silicate clasts. These irons may have formed in melt pools during failed core formation, but are more likely the products of impacts. Many of the ungrouped irons are also probably impact melts, particularly the smaller ones.

Table 3 Here

## 4.2 When and how big did asteroids form?

As discussed in detail below, the formation of chondrites seems to have occurred between roughly 2 and 4 Myr after CAIs (Appendix Question A1). CAIs are the oldest dated objects in meteorites (4,567.3±0.2 Myr; *Connelly et al.*, 2012) and are generally considered to provide the best estimate of the time of Solar System formation, although *Boss et al.,* (2012) have suggested that most CAIs formed toward the end of the Fu Orionis outburst phase of young stellar evolution. Meteorites from differentiated objects seem to have formed slightly earlier than chondrites, although the uncertainties in their ages do allow for significant overlap with chondrite formation and it is possible that some chondrites represent the unmelted crusts of differentiated objects. Some carbonaceous chondrites, such as the CKs, formed roughly contemporaneously with the other chondrite classes, but others may have formed up to 1 Myr later.The absence of chondrites that formed later than 3-4 Myr after CAIs suggests that planetesimal formation effectively ceased at this time in the chondrite forming regions, possibly marking the dissipation of the gas disk. If the carbonaceous chondrites were scattered into the Asteroid Belt from the outer Solar System, this must have occurred after the formation of the youngest carbonaceous chondrites.

At present, there is no way to measure the accretion ages of chondrites directly, although there is evidence from their short-lived $^{53}$Mn-Cr systematics that the H and EH chondrite parent bodies had formed ~2 Myr after CAIs (*Polnau and Lugmair*, 2001; *Shukolyukov and Lugmair*, 2004). In principle, the youngest chondrules in a chondrite group provide an upper limit to the age of their parent body. Most studies have dated individual chondrules using the $^{26}$Al-Mg system on the assumption that the $^{26}$Al/$^{27}$Al ratio of ~5x10$^{-5}$ seen in most undisturbed CAIs represents the homogeneous initial ratio of the Solar System (e.g., *Bizzarro et al*., 2004; *Hutcheon et al*., 2009; *Jacobsen et al.* 2008; *Kita and Ushikubo*, 2012; *Ushikubo et al*., 2013; *Villeneuve et al*., 2009). The assumption of a homogeneous $^{26}$Al/$^{27}$Al ratio in the solar nebula has been questioned by some (*Larsen et al*., 2011; *Schiller et al*., 2015), but at present the balance of evidence supports a homogeneous distribution (*Kita et al.,* 2013; *Wasserburg et al.* 2012). A few chondrule studies have also used the $^{53}$Mn-Cr (*Kita et al*., 2005), the $^{182}$Hf-W (*Kleine et al*., 2008) and the Pb-Pb systems (*Amelin and Krot*, 2007; *Amelin et al*., 2002; *Connelly et al*., 2012; *Krot et al*., 2005) to date chondrules. Together, these studies suggest that: (1) most chondrules in ordinary and CO chondrites and the ungrouped carbonaceous chondrite Acfer 094 formed between 1 and 3 Myr after CAIs, (2) formation of CV chondrules began almost contemporaneously with CAIs and continued for >3.4 Myr, and (3) chondrules in CR chondrites began forming at about the same time as ordinary and CO chondrules, but the youngest ages are >3-4 Myr after CAIs. Chondrules in the CB chondrites, which are different from those in other chondrite groups, have average ages of 4.6±0.5 Myr after CAIs (2σ standard error) and seem to have formed in an impact (*Krot et al*., 2005; *Yamashita et al.,* 2010).

Taken at face value, these chondrule ages imply that all chondrite parent bodies formed >3-4 Myr after CAIs. However, the apparent spreads in chondrule ages are problematic. Chondrules in a particular chondrite group exhibit a limited range or physical and chemical properties that vary from group to group. At the levels of turbulence that are thought to exist in disks, even in a so-called dead zone, chondrules would be mixed over many AU on timescales of 1-3 Myr (*Alexander*, 2005; *Alexander and Ebel*, 2012; *Cuzzi et al*., 2010). Thus, either chondrite formation was dispersed over a large fraction of the solar nebula, as in the Grand Tack and Nice models, or the quoted ranges in chondrule ages are not real. The latter must be ruled out before concluding the former. Indeed, it has been suggested that most of the reported ranges in Al-Mg ages simply reflect the uncertainties in the measurements (i.e., most chondrules in a chondrite group could have the same age or a narrow range of ages) (*Alexander and Ebel*, 2012; *Kita and Ushikubo*, 2012), and that the youngest ages, indeed perhaps all chondrule ages, have been disturbed by parent body processes (*Alexander*, 2005; *Alexander and Ebel*, 2012). It remains to be seen if the range in chondrule Pb-Pb ages also reflect parent body disturbance.

Given that parent body processes can disturb the radiometric systems that are used to date chondrules, it is essential to study only the least metamorphosed and aqueously altered members of a chondrite group. Even in these chondrites, it is essential to take care to select only those chondrules that can be shown to have undergone no secondary modification. The most careful Al-Mg study of chondrule ages conducted to date has been for the ungrouped carbonaceous chondrite Acfer 094 where 9 of the 10 chondrule ages are within error of a mean of $2.3^{+0.5}_{-0.3}$ Myr ($^{26}Al/^{27}Al=5.7\pm1.0x10^{-6}$) after CAIs (Ushikubo et al., 2013). Selecting only similar (type I) chondrules from a previous study of a primitive CO (*Kurahashi et al.*, 2008) gives a mean age of $2.0^{+0.3}_{-0.2}$ Myr ($^{26}Al/^{27}Al=7.1\pm0.8x10^{-6}$) after CAIs. The Al-Mg system is more susceptible to modification in the chondrules analyzed in Semarkona, the ordinary chondrite that has been least affected by parent body processes, but the chondrules have a mean age of $2.0^{+0.5}_{-0.3}$ Myr ($^{26}Al/^{27}Al=7.3\pm1.4x10^{-6}$) after CAIs (Kita et al., 2000; Villeneuve et al., 2009), that is very similar to those for CO and Acfer 094 chondrules. This Semarkona average chondrule age is consistent with an average age of 1.7±0.7 Myr after CAIs for H chondrite chondrules obtained using the $^{182}$Hf-W system (*Kleine et al.,* 2008).

Additional constraints on the timing of accretion can potentially come from the thermal histories of chondrites since the abundance of the radioactive heat source, $^{26}$Al, will have been a function of accretion time. Modeling of mineral ages with different closure temperatures during metamorphism has been used to estimate the accretion time (≥2-3 Myr after CAIs) and size (~100 km in diameter) of the H ordinary chondrite parent body (*Harrison and Grimm*, 2010; *Henke et al*., 2013; *Kleine et al*., 2008; *Trieloff et al*., 2003). Alternatively, assuming that the maximum peak temperatures estimated for any member of a chondrite group represents the peak central temperature achieved in their parent body, along with a diameter of 60 km, *Sugiura and Fujiya* (2014) estimated accretion ages for all chondrite groups (Table 4), including one for the ordinary chondrites of ~2.1 Myr after CAIs. If this accretion age for the ordinary chondrites is approximately correct, it constitutes further evidence that many measured chondrule ages (e.g., those >2.1 Myr) are disturbed. Such estimates depend on many assumptions, not least of which are that we have samples from all depths in the chondrite parent bodies, that the peak temperatures have been accurately determined, that accretion temperatures and initial water ice contents are know, and that the parent body sizes are known. A lack of samples from the deep interior of the CO parent body, for example, could explain why the estimated accretion age of the COs (~2.7 Myr after CAIs) is younger than for the ordinary chondrites despite their average chondrule ages being so similar.

Further complicating these estimates, some have questioned the simple internal heating models and argue that at least the ordinary chondrite parent bodies are rubble piles produced by early collisions while the bodies were still hot (*Ganguly et al*., 2013; Scott et al., 2014). Also, there is the suggestion that the CV chondrites, at least, formed in the presence of magnetic fields generated by dynamos in their planetesimal cores (*Weiss and Elkins-Tanton*, 2013), i.e., that these chondrites are the unheated crusts of differentiated bodies. If true, the accretion age of the CV parent body would have to have been significantly earlier than estimated by *Sugiura and Fujiya* (2014). However, to date, no achondrites or iron meteorites have been linked to the CVs.

Constraints on the timing of aqueous alteration can be established in some chondrites because two products of alteration, fayalitic olivine and carbonates, incorporated short-lived $^{53}$Mn. Fayalitic olivine in ordinary chondrites formed at $2.4^{+1.8}_{-1.3}$ after CAIs, while in CV and CO chondrites it formed at $4.2^{+0.8}_{-0.7}$ and $5.1^{+0.5}_{-0.4}$ after CAIs, respectively (Doyle et al., 2015). 4-5 Myr after CAIs is also about the time that carbonates formed in CM and CI chondrites and the ungrouped chondrite Tagish Lake (*Fujiya et al.,* 2012; *Fujiya et al.,* 2013; *Jilly et al.* 2014). Thermal modeling suggests that the parent bodies of these meteorites formed 3-4 Myr after CAIs.

Thermal modeling of differentiated bodies is even more problematic than for chondrites because, for instance, the initial bulk composition is not known and how the melts segregate within the bodies can have a profound affect on thermal histories (e.g., *Moskovitz & Gaidos* 2011; *Neumann et al*., 2014). Nevertheless, models of varying degrees of sophistication have been used to estimate their accretion times. For instance, core formation on the IIAB, IIAB, IVA, IVB, and IID iron meteorite parent bodies occurred at 0.7±0.3, 1.2±0.3, 1.4±0.5, 2.9±0.5, and 3.1±0.8, Myr after CAIs, respectively (*Kruijer et al.,* 2014). Previous modeling suggested that the accretion of the IVA, IVB, and IID parent bodies occurred, respectively, at 1.0±0.6, 1.3±0.5 and 1.5±0.5 Myr after CAIs (*Kruijer et al.,* 2013). However, *Kruijer et al.,* (2014) have significantly revised the accretion age estimates to ~0.1-0.3 Myr after CAIs for the parent bodies of all these iron meteorites. Estimates for the timing of differentiation and the onset of magma ocean crystallization of the HED parent body (4 Vesta) of 2.5±1.0 Myr (*Lugmair and Shukolyukov*, 1998; *Lugmair and Shukolyukov*, 2001; *Trinquier et al.,* 2008) and $\leq 0.6^{+0.5}_{-0.4}$ (*Schiller et al.,* 2011), respectively, are broadly consistent with estimates of accretion times for Vesta of <1 Myr after CAIs (*Neumann et al*., 2014). Core formation in the angrite parent body occurred ≤2 Myr after CAIs, and its accretion must have occurred ≤1.5 Myr after CAIs (Kleine et al., 2012).

Metallographic cooling rates have been used to argue that the iron meteorites formed in still hot cores that had been stripped of their silicate mantles by impacts (*Yang et al*., 2007, 2008, 2010a,b). The estimated parent body diameters for the best-studied groups are IVA=300±100 km and IVB=140±30 km, and, assuming roughly chondritic compositions, the sizes of the original bodies with their silicate mantles would have been roughly twice these estimates. Thus, the formation of relatively large asteroids began quite early. There is geochemical evidence that an even bigger object, Mars, had reached ~50 % of its present mass in $\leq 1.8^{+0.9}_{-1.0}$ Myr after CAIs (*Dauphas and Pourmand*, 2011; *Tang and Dauphas*, 2014), i.e., contemporaneously with most chondrites.

Table 4 Here

**4.3 How did asteroids form?**

Of the chondrites, only the CI chondrites lack clear evidence for chondrules, although even in these highly altered and brecciated rocks mineral fragments suggest that some chondrules or chondrule fragments may have been present at the time of accretion. It is much harder to say if chondrules were as abundant in the parent bodies of the non-chondritic meteorites, but relict chondrules have been reported in the primitive achondrites, and in silicate inclusions in IIE irons (tentatively linked to H chondrites). Ordinary, R and enstatite chondrites only contain 10-15 vol.% matrix and refractory inclusions are rare in them. It cannot be ruled out that chondrules dominated the dust in their formation regions, but it is also possible that chondrules were preferentially concentrated during planetesimal formation, as predicted by turbulent concentration (*Chambers*, 2010; *Cuzzi et al*., 2010) and possibly streaming instability (*Johansen et al*., this volume) models.

However, the rarity of primitive meteorites with few or no chondrules suggests another intriguing possibility – that chondrule formation and planetesimal formation were linked in someway (*Alexander et al*., 2008). *Metzler* (2012) has argued that chondrules in ordinary chondrites were accreted within hours of their formation while they were still hot and plastic. Such a scenario would be consistent with re-accretion of chondrules produced in hit and run collisions between partially molten planetesimals (*Asphaug et al.*, 2011), for instance. However, such a model must overcome some serious problems: (1) accretion of hot chondrules is difficult to reconcile with the contemporaneous accretion of thermally labile material, such as organic matter and water ice, (2) there are a number of geochemical/petrologic objections to making chondrules this way, and (3) it is not clear why only primitive objects that formed in this way are present in our collections. *Johnson et al*., (2015) also propose that chondrules were produced in collisions, though not rapidly reaccreted, i.e., they are byproducts of planetesimal formation rather than being intimately linked to it. If correct and turbulent concentration and the streaming instability need chondrules to work, then the dominant mechanism(s) for planetesimal formation has still to be identified (Appendix Question A1).

### 4.4 Where did asteroids form?

There is clear isotopic evidence in O, Ti, Cr, Mo and Ru for a distinction between the carbonaceous chondrites and almost all other inner Solar System objects (including Earth, Moon and Mars) for which we have samples (*Burkhardt et al*., 2011; *Warren*, 2011). With the exception of O, the subtle isotopic variations appear to be nucleosynthetic in origin. The causes of these differences are still not understood, but presumably they reflect subtle differences in the materials from which the planets/planetesimals formed (Appendix Question A1). These differences are certainly consistent with formation of the carbonaceous chondrites in the outer Solar System (beyond the initial orbit of Jupiter), as in envisioned by the Grand Tack and Nice models, but at present other explanations may also be possible.

Since objects at or beyond the orbit of Jupiter tend to be water ice-rich, water contents are a potential means for distinguishing between meteorites/asteroids with inner and outer Solar System origins. Some carbonaceous chondrites (CI, CM, and Tagish Lake) were relatively water-rich when they formed, as is evident in the abundant clay minerals that they contain. Other carbonaceous chondrites seem to have accreted some water (CR and CV), but evidence for accretion of water ice in others is limited (CO) or lacking (CK, CH,

CB). Further confusing this issue is that ordinary (*Alexander et al.,* 1989; *Grossman et al*., 2000) and R chondrites (*McCanta et al*., 2008) also seem to have accreted water ice.

If water contents cannot be used to determine whether the carbonaceous chondrite parent bodies formed at greater radial distances than other meteorite parent bodies, perhaps the H isotopes of the water can. There is expected to have been a radial gradient in the D/H ratio of water ice in the disk that reflected mixing between interstellar water (high D/H) and water with low D/H that had re-equilibrated with $H_2$ in the hot inner disk (*Albertsson et al*., 2014; *Jacquet and Robert*, 2013; *Yang et al.,* 2013). The D/H ratios of water in all comets are similar to or greater than the terrestrial ratio (*Lis et al*., 2013), and the terrestrial ratio is almost an order of magnitude higher than the bulk solar value. The only other outer Solar System planetesimals with measured D/H ratios are Saturn's moons Enceladus, whose water like most comets is enriched in D relative to the Earth (*Waite Jr et al*., 2009), and Titan, whose atmospheric methane D/H is Earth-like (*Nixon et al.*, 2012). On the other hand, the N isotopes of Titan's atmosphere are $^{15}$N-enriched compared to the Earth (*Mandt et al*., 2014), like comets (*Rousselot et al*., 2014; *Shinnaka et al.*, 2014). While it is still not certain where and how the comets and Saturn's moons formed, their compositions suggest that ice in the outer Solar System had D/H ratios that were greater than or equal to terrestrial. However, except for the CR chondrites, the best estimates of the D/H ratios of water in carbonaceous chondrites at the time of accretion were sub-terrestrial (*Alexander et al.,* 2012). Rather surprisingly, the D/H ratios for water in ordinary and R chondrites are similar to those of the more enriched comets. This probably does not mean that they formed in the outer Solar System. It more likely reflects isotopic fractionation associated with the loss of $H_2$ generated by the oxidation of metal by water in their parent bodies. The effect may have been larger in the ordinary and R chondrites because they had lower initial water/metal ratios. Nevertheless, this process should also have affected the carbonaceous chondrites, so that their initial water D/H ratios were probably somewhat lower than estimated by *Alexander et al.,* (2012).

In Summary, while carbonaceous chondrites and their asteroidal parent bodies are geochemically distinct from other inner Solar System objects that we have samples of, at present there is no direct evidence that they formed at much greater radial distances from the Sun than other meteorite parent bodies.

## 5. Asteroid Compositional Trends

In this Section, we explore questions related to the taxonomic distribution of asteroids at both the large and small sizes. A number of questions listed in the Appendix are discussed here.

### 5.1 What is the source of the compositional gradient in the Main Belt?

For the largest asteroids (D>~100km), the trend of taxonomic types as a function of distance from the Sun, in the order of S, C, P, D, remains robust. The canonical interpretation of this gradient is that it reflects the compositional variation with distance from the Sun of the dust in the solar nebula (*Chapman et al., 1971, 1975, Gradie & Tedesco 1982,* Question C1 Appendix). At the time, it was anticipated that (1) dynamical processes might have smeared an originally sharper division between asteroids of different composition, and (2) that different materials, condensing and accreting earlier or later, might have formed at the same solar distance, also smearing the gradient (*Ruzmaikina et al., 1989, Petit et al., 2002*).

These ideas have not been ruled out entirely, but it is now apparent that other processes may have played a greater role in producing this gradient. The largest source of uncertainty is the ongoing challenge to associate many of the asteroid taxonomic types with specific types of meteorites or compositions (*Burbine et al.,* 2002*).* There are ~40,000 meteorites in our collections, yet only ~100-150 distinct parent bodies have been identified from the collections. There are over 700,000 known asteroids, but only ~100 asteroids with diameters greater than 100 km and 122 notable families (see *Nesvorny et al.,* this volume) that may be considered indicators of the number of parent bodies existing in the Main Belt today (Appendix Question B1). It is also possible, but not yet firmly established, that more than one asteroid could be the source of a single meteorite group (e.g., *Vernazza et al.,* 2014; Appendix Question B2). There is certainly more work needed to understand the links between meteorites and their asteroidal sources. Next, we briefly summarize some of the known and proposed associations.

It is now clear that a fraction of S-complex objects are ordinary chondrites and that the HEDs are mostly associated with Vesta and its family, but most taxonomic types cannot be robustly associated with specific meteorites or mineralogies. Many asteroid spectra are simply not diagnostic of unique mineral assemblages. E-types have been associated with enstatite chondrites or achondrites, M-types with metallic meteorites, A-types with pallasites or olivine-rich achondrites, L-types with CAIs, etc., but these connections are not unique or robust (*Burbine et al.,* 2002). It is generally understood that some of the C-complex asteroids (including P- and D-types) are likely associated with carbonaceous chondrites, of which there are many subtypes. It is probable that many of these asteroids do not supply meteorites to Earth because their distant orbits are less likely to be perturbed into Earth-crossing, or if their fragments do enter the atmosphere, they may be too fragile and/or arrive at too high a velocity to survive as a meteorite.

A major advance toward interpreting the taxonomic gradient is the understanding of processes that move asteroids and their fragments about. Since the time of *Asteroids I and II*, there is far better understanding of major and minor resonances in the Main Belt (*Nesvorny et al*., 2002), of how the Yarkovsky effect (and associated YORP effect) works (*Bottke et al., 2006*), of past dynamical regimes (e.g., the Grand Tack, or Nice models) that may have had profound effects on the distributions of small bodies in the inner and outer Solar System (*Morbidelli et al., Chapter*), and of how asteroid families produced by catastrophic collisions evolve. It has also been suggested that some Inner Belt asteroids were emplaced as debris from the terrestrial planet region (*Bottke et al.,* 2006), that comets can be implanted into the Main Belt, and even that Ceres might have been implanted from the outer Solar System (*McKinnon, 2008*).

The gradient is less obvious for smaller Main Belt asteroids than for larger ones, but this could be attributed to the fact that several processes act more readily on smaller bodies. The idea that the larger asteroids retain some element of a primordial gradient is more tenable, but clearly only one of a number of possibilities, given recent models (*Walsh et al.,* 2011, 2012). It is uncertain whether astronomical observations together with dynamical scenarios can uniquely solve this problem. We need to measure actual compositions as a function of location in the Solar System, whether through better understanding from meteoritical research or from in situ studies, to resolve the matter. It is still not understood in any quantitative way how meteorites with different oxygen isotopes were distributed in the early Solar System, nor from where they are derived in the modern epoch. Short of a wholly impractical mineralogical and isotopic assay of dozens of asteroids and comets, there does not appear to be a way for this question to be robustly resolved in the near term.

**5.2 Distribution Changes with Size**

While the smallest asteroids are the least studied due to observational biases, there are some interesting possible trends and discoveries that have been noted in the past decade (Appendix Question C2). Among the Hildas and Trojans the relative frequency of D- and P-types changes as a function of size. *Grav et al.,* (2012) find a decrease in spectral slope with decreasing size, which translates to more P-types at smaller sizes. *DeMeo & Carry* (2014) also note a change in relative fraction between the D- and P-types. The cause of this change in spectral slope as a function of size is not yet understood. Possible explanations could include space weathering (*Lantz et al.,* 2013, *Brunetto et al.,* 2014), compositional differences, or grain size effects (e.g., *Cloutis et al.,* 2011a,b).

Additionally, among the smallest asteroids (D <~15 km) we find very red asteroids, D-types, in the Inner Belt where they were previously not expected (*DeMeo et al., 2014*). The compositional make up of these bodies and how they arrived in their current locations are under debate. They could have been scattered farther than expected during the late-stage migration modeled by *Levison et al.,* (2009). Perhaps they arrived through another mechanism such as an earlier migration, other planetary scattering, or Yarkovsky drift across the resonances. Ultimately, we cannot rule out the possibility that they are compositionally distinct from other D-types and thus do not require an implantation mechanism.

*DeMeo & Carry* (2013, 2014) reported that the relative abundance of C-complex asteroids in the Inner Main Belt is greater at the smaller sizes. There are multiple possible explanations for asteroids of different compositional types to have different size distributions. It could be related to the size-frequency distribution of families, for example C-complex objects may survive longer if they have high porosity, as in the case of Mathilde. Bodies with an ice-component may have shorter lifetimes as both sublimation and collisions are at play (*Rivkin et al*., 2014). Shocked S-complex objects that have suppressed absorption features could be masquerading as other taxonomic types, including the C- or X-complex (*Britt et al.,* 1989, *Reddy et al.,* 2014, *Kohout et al.*, 2014). Differences could also be attributed to the primordial size distributions of asteroids of different compositional types.

**5.3 What processes can mix wildly different meteorite types into a single tiny body?**

There has been increased focus recently on asteroids in the size range for meteorite delivery, partly due to recent spectacular events such as the fall of 2008 $TC_3$ (Almahata Sitta), the airblast from the Chelyabinsk meteorite in 2013, and several asteroid close approaches, such as 2012 $DA_{14}$.

Almahata Sitta presents a new puzzle to the meteorite and asteroid communities because of the diversity of meteorite types that have been mixed into a single small body (Appendix Question B6). Almahata Sitta is predominantly made of ureilitic lithologies, but contains an unprecedentedly large fraction of a variety of other meteorite types, including enstatite chondrites, ordinary chondrites, and even carbonaceous chondrites (*Zolensky et al*., 2010). *Gayon-Markt et al.,* (2012) concluded that it could not have been assembled by low velocity collisions among asteroids of diverse types in the Nysa-Polana family region (the suspected source region of Almahata Sitta) in the current dynamical environment. The mixture of two or three different meteorite types could be ascribed to unusual but possible coincidental encounters at the low-velocity tails of their dynamical parameters, but the variety of different materials in this case is so large that it is difficult to appeal to such low-probability events. In addition, Almahata Sitta is not a regolith breccia (it does not contain implanted solar-wind particles), so it was not lying on the surface of an asteroid for an extended period of time. Hence, *Gayon-Markt et al*., (2012) suggest that the diverse pieces must have been assembled in early times when collisional velocities were much lower.

Asteroid 2008 $TC_3$ exhibited a seemingly flat reflectance spectrum most closely, but not convincingly resembling F-types (*Jenniskens et al.,* 2009; F-types are from the Tholen taxonomy and would fall under the C-complex in the Bus and Bus-DeMeo taxonomies, see Table 1), where the spectral data had large error bars and did not contain diagnostic absorption features. Ureilites are fairly dark and some of the other incorporated meteorite types exhibited shock-darkening, thus a fairly featureless reflectance spectrum is not unexpected. A wide variety of mineralogies, so long as they contain a fraction of low albedo materials or shock-darkened grains, might show such a spectrum. Thus there is no strong basis for arguing that other asteroids with similar spectra should be similarly composed of heterogeneous materials the way Almahata Sitta is.

This mystery might be addressed if we could identify a larger proportion of meteorites that resemble Almahata Sitta. Is there a bias against finding meteorites of this type? The Sudanese desert was an especially good background for picking up pieces of the meteorite. Maybe these kinds of meteorite conglomerates are typically even weaker than Almahata Sitta and fragment even higher in the atmosphere, lessening the chance of recovery on the ground (e.g., similar to reasons for why we apparently lack samples of comets). If this is a more common kind of assemblage than we now think, we would be forced to abandon some of our ideas about asteroid family homogeneity (e.g., from the similar colors and albedos of asteroid family members) or about relative velocities among asteroids during early epochs of Solar System history. Given the mismatch in scale between astronomical observation of asteroids and the centimeter-to-meter-scale heterogeneities within Almahata Sitta and other meteorites such as Kaidun (*Zolensky* 2003), asteroid astronomical techniques are less likely to resolve this mystery than are techniques of meteor and fireball observation and meteorite research in the laboratory.

**5.4 The Missing Mantle Problem**

The canonical view of asteroid differentiation has been that a fully differentiated body should form an iron core, a silicate mantle, and a basalt crust. Olivine and pyroxene are the spectrally dominant materials in the mantle and crust, respectively. Meteorite collections include a diversity of iron meteorites that imply the existence of over 60 distinct parent bodies (*Wasson* 1995, *Burbine et al.*, 2002). However, the meteorite and asteroid records lack a substantial population of olivine-dominated or pyroxene-dominated bodies (*Burbine et al.,* 1996, *Bus & Binzel 2002a*, *Binzel et al.*, 2004, *Lazzaro et al.*, 2004, *Carvano et al.,* 2010, *DeMeo & Carry* 2013), aside from HEDs and Vestoids linked to Vesta (*Consolmagno & Drake* 1977, *Binzel et al.*, 1993, *Prettyman et al.,* 2012). This has been known as the Missing Mantle Problem, and also as the Missing Dunite Problem (Appendix Question C3). One of the major outstanding questions in asteroid science existing for decades has been this perceived shortage of dunite or olivine in the Asteroid Belt.

Multiple solutions have been proposed to the missing mantle problem. The "battered to bits" scenario was a leading theory from the 1990s, whereby *Burbine et al.,* (1996) further explored a proposal by *Chapman* (1986) that the mantle and crustal material of the original differentiated bodies had been ground down to pieces below the limit of our observational capabilities or even so they were too small to be delivered to Earth as meteorites. The primary weakness of this scenario is that the mantle and crustal components should not be substantially structurally weaker than undifferentiated material such as the C- and S-complexes that have survived. Additionally, larger scale spectroscopic surveys have placed even more stringent limits on the abundance of differentiated material and find it exceedingly rare in both the Main Belt and among NEOs that are sourced from the smaller (m- to km-sized) Main Belt population (*Bus & Binzel 2002a*, *Binzel et al.*, 2004, *Lazzaro et al.*, 2004, *Carvano et al.,* 2010, *DeMeo & Carry* 2013).

A more recent suggestion is that the classic view of asteroids differentiating into a pyroxene-rich crust, olivine-rich mantle, and iron core may be incorrect or uncommon (*Elkins-Tanton et al., 2011*). The differentiation models cause an asteroid's interior to heat and melt due to the decay of $^{26}$Al, but the exterior remains unheated and primitive (see chapter by *Scheinberg et al.*). Thus the unheated crust would hide evidence of a differentiated interior from a remote observer. *Weiss et al.* (2012) propose that asteroid Lutetia's high density, measured by the Rosetta spacecraft, is evidence of a core due to internal differentiation. Also, recent measurements of the primitive CV meteorite Allende find a directionally stable remnant magnetization that could be explained by the presence of a core dynamo in the parent body's past (*Carporzen et al.,* 2011). As for rethinking the olivine-rich mantle model, the smoking gun has been the lack of evidence for olivine on the surface of Vesta from the Dawn mission (*Le Corre et al.*, 2013). Another test of this differentiation scenario could be observations of asteroid families, whereby family members represent pieces of the parent asteroid's interior. The spectra and albedos of asteroid families tend to be very homogeneous (*Parker et al., 2008, Masiero et al., 2011*), however, the parent bodies of these families may not have been large enough to differentiate in the first place.

In addition to improved differentiation models, there is a new dynamical context for small bodies. It is possible, even likely, that bodies that formed iron cores formed closer to the Sun, in the terrestrial planet region (see chapter by *Scott*). During the tumultuous time of planet formation, these differentiated bodies were catastrophically disrupted, but their remnants were later scattered outward into their current locations in the Main Belt (*Bottke et al.*, 2006). In addition, the so-called Grand Tack models (*Walsh* *et al., 2011*) proposes that within the first few million years of Solar System formation Jupiter migrated inward to the current location of Mars, then due to Saturn's presence turned course and moved back outward, scattering a significant number of bodies out to the location of the Main Belt today. The orbital distribution of basaltic and olivine-dominated objects spread across the Main Belt today are consistent with the idea that a small amount of material was scattered into their current locations (e.g., *Moskovitz et al., 2008*).

**5.5 The Ordinary Chondrite Paradox**

The ordinary chondrite paradox was a longstanding puzzle in asteroid science (Appendix Question B7), whereby the most common meteorite type, the ordinary chondrites, did not spectroscopically match the most common asteroids, the S-complex (*Bell* 1989). Originally the S-complex asteroids in the Inner Belt were interpreted as igneous bodies that had undergone significant heating and melting, exposing iron, olivine, and pyroxene on the surface (*McCord & Gaffey* 1974, *Chapman* 1974). Thus, they were not considered compositionally linked to ordinary chondrites. However, the possibility of an effect, such as space weathering seen on the Moon, that altered ordinary chondrite (OC) spectra to look like S-complex asteroids could not be ruled out (*Chapman* 1979).

By the 1990s, the space weathering hypothesis gained traction. Spectral measurements of tens of NEOs demonstrated that the S-complex NEOs spanned the entire spectral slope range from spectrally flat OC meteorites to spectrally red S-complex Main Belt asteroids (*Binzel et al.,* 1996, 2010). The Galileo mission data for Ida revealed that spectral differences were correlated with the age of surface features, and the discovery of Ida's moon Dactyl allowed density estimates that rejected the hypothesis of a core or other large metal components that may have been produced by large-scale melting and differentiation (*Chapman* 1996). Images and spectra of asteroid (433) Eros from the NEAR mission showed that dark regolith on the wall of the crater Psyche moves downslope exposing younger, brighter material (*Veverka et al.,* 1999; *Murchie et al.,* 2002). The most important result from NEAR-Shoemaker was the conclusion from x-ray fluorescence spectroscopy that the major elemental composition of Eros was that of ordinary chondrites and that body had not differentiated (*Nittler et al.,* 2001). Additional ground-based studies of Itokawa (*Binzel et al.*, 2001) and an increased understanding of the effects of space weathering (*Sasaki et al.,* 2001, *Chapman* 2004, *Brunetto et al.,* 2006) further strengthened the OC-S-complex link. However, it was not until the sample return from Itokawa by JAXA's Hayabusa spacecraft (*Nakamura et al.,* 2011) that conclusively proved the link. While it became clear that space weathering accounted for much of the spectral slope difference between OC meteorites and S-complex asteroids, other factors including grain size and observational phase angle were also determined to affect spectral slope (see chapter by *Reddy et al.*). For further discussion of the ordinary chondrite problem see the chapter by *Brunetto et al.*

The ordinary chondrite paradox case is now generally considered closed - ordinary chondrite compositions are part of the S-complex. There are two large problems remaining, to determine (1) which S-complex asteroids are not OC's and may supply other meteorite types, and (2) how many original parent bodies the OC's come from. The S-complex encompasses a range of spectral characteristics and a diverse set of compositions (e.g., *Gaffey et al.,* 1993, *Dunn et al.,* 2013). Detailed mineralogical modeling of a large quantity of S-complex spectra, combined with a refined understanding of other factors that affect an asteroids reflectance spectrum, is needed if we are to make further progress. Exploring a broader wavelength range from the UV to mid-infrared would provide additional constraints.

## 6. Asteroids in their greater context

Studies of asteroids and meteorites reveal their current compositions and structure, how and when they formed including the conditions of the Solar System at the time of planet and planetesimal formation. Asteroids, however, are more broadly relevant in our own Solar System and beyond. In this Section we explore asteroids and their relevance to Earth and habitability and the topic of asteroids outside our Solar System. The field of exo-asteroids is growing rapidly and the role asteroids play in creating the conditions for life (particularly the role in delivering water) is now being studied extensively. While these topics play a minor role in this book, one may envision these topics to be mature by the time of *Asteroids V*.

### 6.1 Asteroids relevance to Earth formation and conditions for life

No meteorite group or combination of groups can reproduce both the elemental and isotopic compositions of the Earth, the planet for which we have the best data (*Palme and O'Neill*, 2014; *Halliday*, 2014: Appendix Question D1). Impact erosion and volatile loss during melting/differentiation of planetesimal precursors may explain some of the differences, but it seems that the Earth was at least partially made of materials not present in our meteorite collections.

The abundances of highly siderophile (metal-loving) elements in the Earth's silicate mantle and crust are much higher than would be predicted since these elements should have gone essentially quantitatively into the Earth's Fe-metal core during differentiation (*Walker*, 2009). It has been proposed that these elements were added to the Earth after the last major event that enabled exchange between the core and the silicate mantle (*Chou*, 1978), which is normally taken to be the Moon-forming impact. The age of the Moon-forming impact is still uncertain – it must have been earlier than the oldest known lunar crustal rocks that formed 4.36 Byr ago (*Borg et al.*, 2011), and could be as old as $4.507^{-0.1}_{+0.01}$ Byr (*Kleine et al.*, 2009; *Touboul et al.,* 2007). The isotopic composition of the highly siderophile elements Os and Ru suggest that the material responsible for this so-called late veneer was probably ordinary- or enstatite-chondrite-like (i.e., volatile-poor), rather than carbonaceous-chondrite-like (i.e., volatile-rich) (*Walker*, 2009). The fraction of the Earth's mass that was accreted as a late veneer was of the order of 0.5 % (*Walker,* 2009). The material accreted during the proposed late heavy bombardment (LHB), an increase in the lunar basin formation rate about 4 Byr ago, would have been a minor component of the late veneer. However, there has been a longstanding debate about how long the LHB lasted or even whether there was a LHB (*Chapman et al.,* 2007; *Morbidelli et al.,* 2012). One model (the Nice model) proposes that the LHB was caused by an episode of giant planet orbital migration that destabilized the inner edge of the Asteroid Belt (*Bottke et al.,* 2012).

The LHB, and the late veneer as a whole, was not a significant source of the Earth's volatiles. Apart from the Os and Ru evidence, and model support that the LHB and the late veneer material were volatile-poor, there is evidence for oceans 4.4-4.3 Byr ago (*Cavosie et al.,* 2005; *Mojzsis et al.,* 2001; *Wilde et al.,* 2001), i.e., 300-400 Myr before the LHB. The Earth's atmosphere is probably even older, ~$4.527^{+20}_{-10}$ Byr (*Avice and Marty,* 2014), which is of a similar age to or possibly older than the Moon. So volatiles were accreted by the Earth quite early and were not lost in any subsequent giant impacts, which is consistent with the new evidence for a wetter Moon than previously suspected (*Saal et al*., 2013) and with the timing for water accretion measured in eucrites (*Sarafian et al*., 2014).

It has been argued that Jupiter family comets (JFCs) like Hartley 2 could have been the main sources of the Earth's water because they have roughly terrestrial water D/H ratios (*Hartogh et al.,* 2011), although the water D/H of another JFC is roughly three times terrestrial (*Altwegg et al.,* 2015). However, at least in the Nice model objects from the scattered disk, the source of the JFCs, would have been scattered into the inner Solar System only during the LHB (*Levison et al.,* 2009), which is too late given the evidence for an atmosphere and oceans 100s of millions of years before the LHB. Also, as mentioned above, the isotopic compositions of Os and Ru suggest that the late veneer material was dominated by ordinary- and enstatite-chondrite-like material. Finally, it is important to remember that the Earth would have accreted whole comets and not just their ice. Comets are organic-rich and judging by the organics in meteorites and interplanetary dust particles, which may be cometary, cometary organic material is very D-rich. Consequently, the bulk D/H even of comets like Hartley 2 are probably significantly more D-rich than their ice and the Earth (*Alexander et al*., 2012). On the other hand, the bulk H and N isotopic compositions of CI and CM chondrites are quite similar to the Earths (*Alexander et al*., 2012). In fact, the addition of ~2 wt.% of CI or ~4 wt.% of CM chondrite material can roughly reproduce not only the H and N isotopes, but also the abundances of many of the Earth's most volatile elements (H, C, Cl, Br, I, Ne, Ar and Kr) (*Alexander et al*., 2012; *Marty*, 2012).

### 6.2 Asteroids outside our Solar System

Asteroids play an important role outside of our Solar System as well (Appendix Questions D3 & D4). Asteroids are inferred to exist as remnants around white dwarf stars and are used as tracers of the dynamics and physical properties in these systems. Gravitational settling in cool white dwarfs causes all elements heavier than helium to sink to the star's interior. One fourth to one third of cool white dwarfs have heavy elements in their atmospheres that accreted from in-falling asteroids (*Alcock et al., 1986, Jura 2008, Jura & Young 2014*). These asteroids around white dwarfs are at least 85% by mass oxygen, magnesium, silicate, and iron, resembling rocky planetesimals and bulk Earth in our Solar System (*Jura & Young 2014*). This observational technique of studying white dwarf atmospheres can also detect asteroid differentiation and evidence of water was detected in "exoasteroid" debris (*Farihi et al., 2011, 2013*). Studies of cool white dwarfs may enable us to understand asteroid compositions, formation and evolution in other planetary systems.

Asteroid Belts are present around stars, as evidenced by the collisional dust (equivalent to our own zodiacal dust) observed around other mature stars. Debris disks are detected by the light emitted by dust grains in the disk at thermal wavelengths (e.g. *Lawler et al., 2009, Lawler & Gladman 2012*). The peak flux and wavelength of the emitted light provides an estimate for the mass of the disk and its distance from the star. While most known debris disks are Kuiper-Belt-like, located at distances from tens to hundreds of AU, a small number exhibit mid-infrared emission indicating warm dust. Most spectral measurements of debris disks have been taken at infrared wavelengths where the flux ratio of the disk to star is most favorable. Minerals can be identified by emission features (e.g., *Christensen et al., 2000*) with silicates being the most common and easiest to identify (e.g., *Beichman et al., 2005, Lisse et al., 2012, Ballering et al., 2014*). Interpretation of these spectra, however, is challenging because many grain properties other than composition (size, shape) can affect the spectrum. Continued studies of debris disks will reveal the diversity of extrasolar asteroid belt configurations and compositions that will provide context for our own Solar System and for understanding the role asteroid belts play in creating the conditions for habitability on Earth-size planets.

**7. The Future: What's next?**

Potential avenues for future work on specific questions addressed in this chapter are given in Table 5. We highlight a few avenues for future work here:

- In general, for the future of understanding the compositions and distributions of asteroids there are several techniques and topics that hold promise. While the past decades have focused primarily on the visible and near-infrared wavelength ranges for spectroscopic study, extending to the UV and mid-IR could potentially provide new constraints for asteroid surfaces.
- The internal structures and geophysics of asteroids remain largely unexplored. Additional studies of densities and differentiation processes will play an important role in understanding asteroid interiors (*Carry* 2012).
- Studies of small asteroids (a few km in the Main Belt and a few tens of meters among NEOs) will be a focus in future years as there is much progress to make on the processes relevant to these size ranges. Understanding the compositional makeup and mixture of serendipitous meteorite falls is a young and developing field. The study of shock darkening effects on the surfaces of asteroids and meteorites is another avenue of future work. While the foundations for the shocking process were established decades ago (e.g., *Britt & Pieters* 1994), there is renewed interest as we observe the smallest asteroids.
- Ongoing and upcoming space missions will also add great value to our understanding of asteroid compositions. Dawn recently completed its study of Vesta and will have studied Ceres by the time this book is published. OSIRIS-Rex and Hayabusa 2 are scheduled to return samples of primitive asteroids. Gaia is actively taking visible wavelength spectra of tens to hundreds of thousands of asteroids and will provide a large new set of data to the community.

Just as important as exploring all these interesting avenues of asteroid research is exploring the connections between them. We must understand how each new discovery relates to our current body of knowledge and to the big picture of asteroids, our Solar System, and beyond.

**FIGURES**

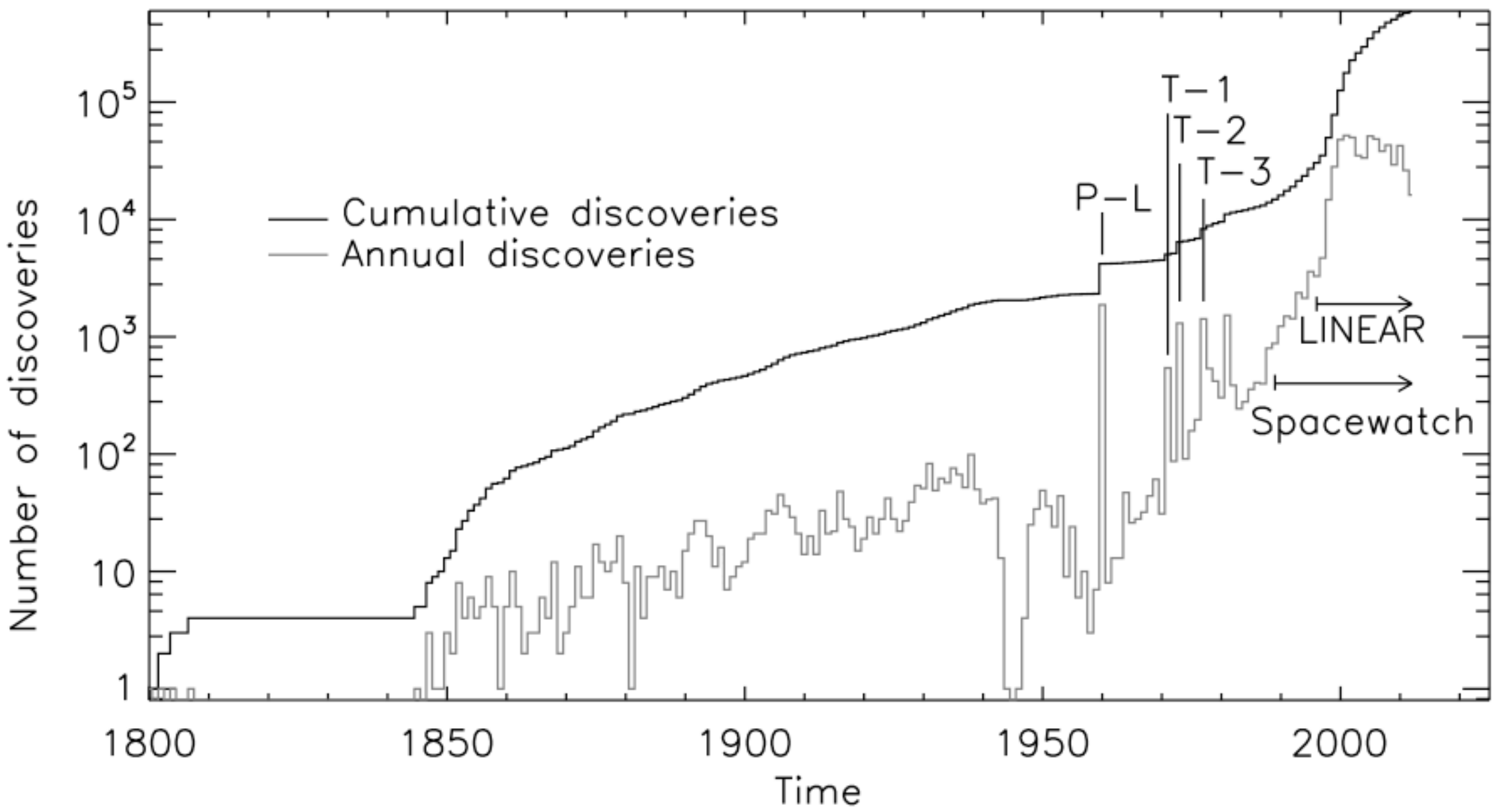

Fig 1. The cumulative number of known asteroids and the yearly discovery rate are plotted above. The surveys responsible for spikes in detection are marked, the Palomar-Leiden (P-L), Trojan (T-1, T-2, T-3), Spacewatch, and LINEAR. Recent years have seen an explosion in asteroid discoveries due to automated telescopic surveys with advanced detection algorithms. By the year 2015, nearly 700,000 asteroids have been discovered.

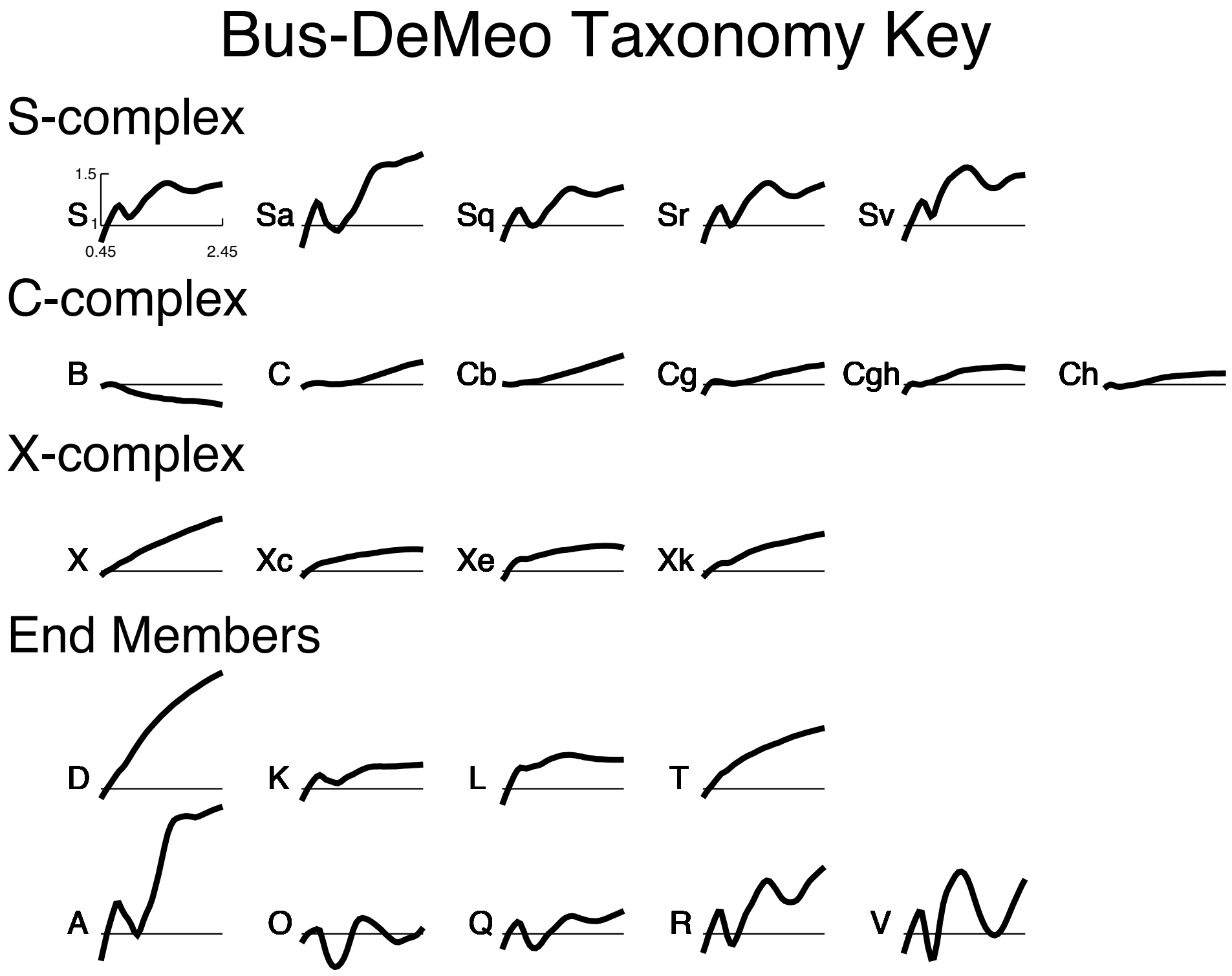

Fig 2. Asteroid Taxonomy Key. The 24 spectral classes of the Bus-DeMeo taxonomy key measured over visible and near-infrared wavelengths. Based on work from *DeMeo et al.*, 2009.

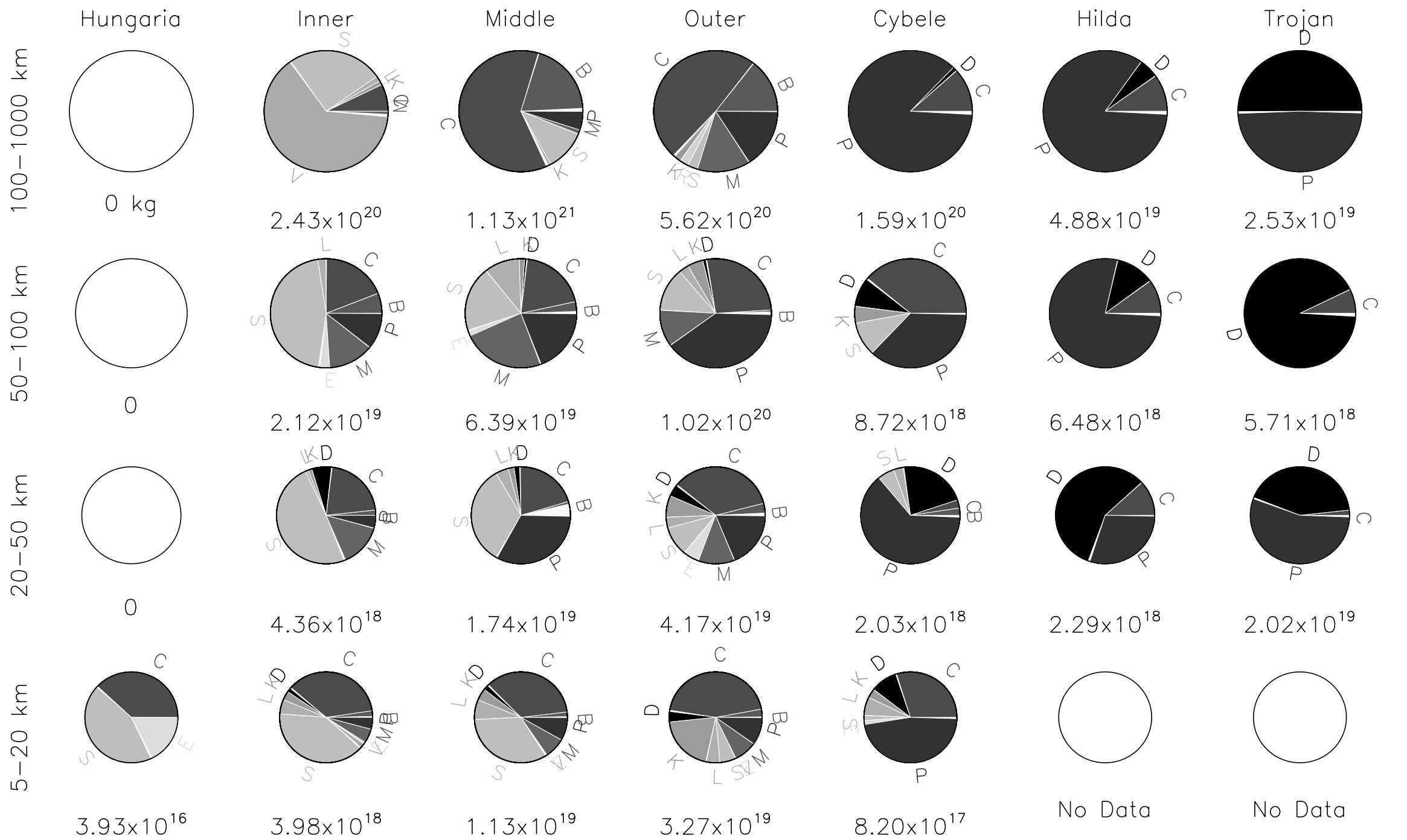

Fig 3. The distribution of asteroid classes by mass in distinct size ranges and distances from the sun. Asteroid mass is grouped according to objects within four size ranges, with diameters of 100–1,000 km, 50–100 km, 20–50 km and 5–20 km. Seven zones are defined as in Fig. 1: Hungaria, inner belt, middle belt, outer belt, Cybele, Hilda and Trojan. The total mass of each zone at each size is labeled and the pie charts mark the fractional mass contribution of each unique spectral class of asteroid. The total mass of Hildas and Trojans are underestimated because of discovery incompleteness. The top row is consistent with results from *Gradie & Tedesco* 1982 and *Gradie et al*., 1989. The figure is from *DeMeo & Carry* 2014.

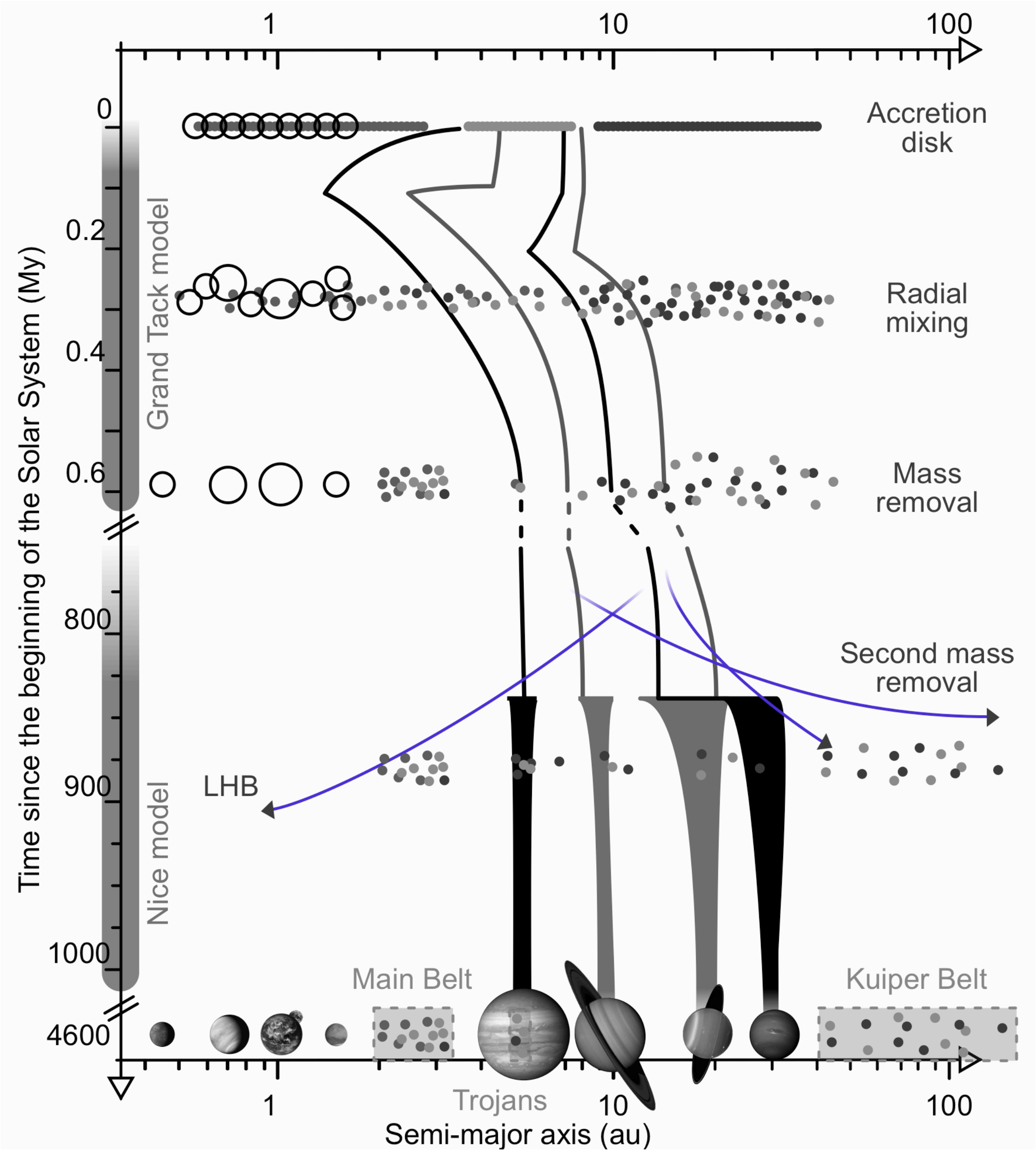

Fig 4. This cartoon depicts major components of the dynamical history of small bodies in the Solar System based on models. These models may not represent the actual history of the Solar System, but are possible histories. The figure displays periods of radial mixing, mass removal and planet migration— ultimately arriving at the current distribution of planets and small-body populations. This figure is from *DeMeo & Carry* 2014.

TABLES:

Table 1: Comparison and explanation of the progression of spectral taxonomies. A simple description of relevant minerals and meteorites for each class is presented. For meteorite and mineral details see *Burbine et al.*, 2002.

# Comparison of Spectral Taxonomies

| Taxonomic System: | Tholen (1984) | Gaffey (1993) | Bus (2002) | Bus-DeMeo (2009) | Taxonomy Notes | Relevant minerals possible meteorite analogs (for more details see Burbine et al., Asteroids III) |
|---|---|---|---|---|---|---|
| Wavelength Range: | 0.33-1.1 um | 0.35-2.50 um | 0.45-0.90 um | 0.45-2.45 um | | |
| S-Complex | 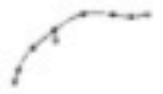 S | SI<br>SII<br>SIII<br>SIV<br>SV<br>SVI<br>SVII | S<br>Sa<br>Sq<br>Sr<br>Sk<br>Sl | 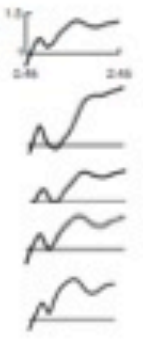 S<br>Sa<br>Sq<br>Sr<br>Sv | **Tholen:** Defined only S.<br>**Gaffey:** 7 mineralogic classes based on Band I center & Band II / Band I area ratio. Primarily separates olivine to orthopyroxene ratio.<br>**Bus:** Separates based on strength of 0.9um drop, indicative of 1um band.<br>**B-D:** Definition largely preserved from Bus. Now includes full 1um feature and 2 um feature in near-ir. Sl & Sk are removed, Sv is added. | **Minerals:** olivine,pyroxene<br><br>**Meteorites:** S(I): Pallasites?, R chondrites, Brachinites, S(IV): many are ordinary chondrite-like, S(V): Primitive achondrites?, S(VII): Basaltic Achondrites |
| C-Complex | 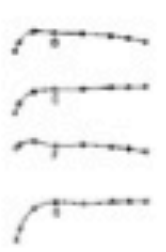 B<br>C<br>F<br>G | | B<br>C<br>Cb<br>Cg<br>Cgh<br>Ch | 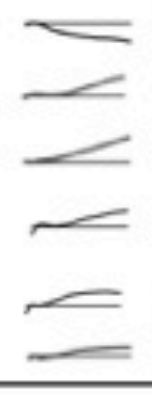 B<br>C<br>Cb<br>Cg<br>Cgh<br>Ch | **Tholen:** Primarily distinguished by the 0.3-0.5um UV dropoff region. Bus & B-D do not cover this region, thus do not make these distinctions<br>**Bus:** Defined by UV dropoff and/ or by 0.7um Cgh, Ch feature.<br>**B-D:** Definition largely preserved from Bus. Near-infrared is largely degenerate. | **Minerals:** opaques, carbon, phyllosilicates, some have weak features indicating olivine, pyroxene<br>**Meteorites:** carbonaceous chondrites (except CV) , possibly impact melts from ordinary chondrites and HEDs? |
| X-Complex | 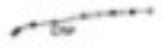 E<br>M<br>P | | X<br>Xc<br>Xe<br>Xk | 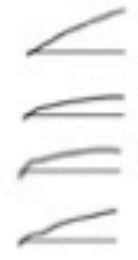 X<br>Xc<br>Xe<br>Xk | **Tholen:** EMP are spectrally degenerate. Distinguished by high (E), med (M) and low (P) albedo.<br>**Bus:** X class defined by shape of curve and/or 0.49um Xe feature.<br>**B-D:** Definition largely preserved from Bus. Near-infrared is largely degenerate. | **Minerals:** M,P: opaques, carbon, low-Fe pyroxene. E: enstatite, oldhamite<br>**Meteorites:** M,P: carbonaceous chondrites (not CV). M w/high radar albedo: irons, CB condrites, Silicate rich irons. E: enstatites |
| Other: End Members, Outliers | 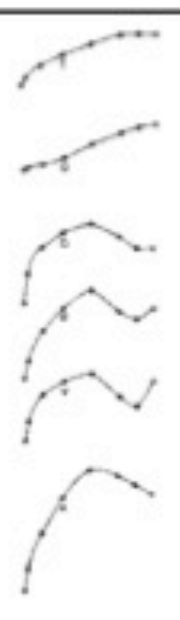 T<br>D<br>O<br>R<br>V<br>A | | T<br>D<br>Q<br>O<br>R<br>V<br>A<br>K<br>L | 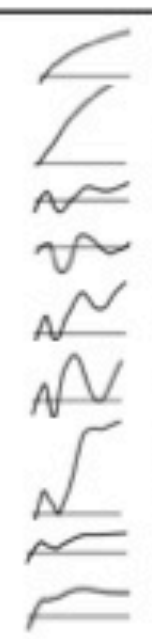 T<br>D<br>Q<br>O<br>R<br>V<br>A<br>K<br>L | Definitions for each of these classes are relatively consistent among taxonomies as they are each spectrally distinct. | D opaques, organics<br>Q mostly LL OCs<br>O pyroxene, olivine<br>R olivine, pyroxene<br>V HEDs, pyroxene, plagioclase feldspar<br>A pallasite, brachinite, R chondrites, olivine.<br>K CO,CV. olivine<br>L CAI-rich, spinel-rich |

Table 2. Summary of the recognized chondrite classes and groups.

| Class | Groups | |
|---|---|---|
| Enstatite (E) | EH, EL | Highly reduced, metamorphosed types 3-6 |
| Rumuruti (R) | | Metamorphosed types 3-6, some OH-bearing minerals |
| Ordinary (O) | H, L, LL | Metamorphosed types 3-6, evidence for $H_2O$ activity. |
| Carbonaceous (C) | | |
| | CI, CM, CR | Aqueously altered types 1-3 |
| | CO, CV | Metamorphosed types 3, evidence for $H_2O$ activity. |
| | CK | Metamorphosed types 3-6, evidence for $H_2O$ activity. |
| | CH, CB | Metal-rich, impact products |
| Kakangari (K) | | |

Table 3. Summary of the non-chondrites.

| | |
|---|---|
| **Primitive achondrites** | |
| Winonaites | Related to IAB irons |
| Acapulcoites and lodranites | |
| Brachinites | |
| | |
| **Achondrites** | |
| Aubrites | Broadly basaltic compositions, related to E chondrites |
| Ureilites | Contain veins of graphite and diamond |
| Angrites | Basaltic compositions |
| HEDs | From 4 Vesta |
| *Howardites* | Eucrite and diogenite breccias |
| *Eucrites*[1] | Basaltic compositions |
| *Diogenites* | Coarse pyroxene and olivine |
| | |
| **Stony irons** | |
| Mesosiderites | Impact breccias |
| Pallasites | Core-mantle boundary material (?) |
| *Main group* | Related to IIIAB irons (?) |
| *Eagle Station group* | Related to C chondrites |
| *Pyroxene group* | |
| | |
| **Irons**[2] | |
| Magmatic | IVAs related to L/LL chondrites |
| *IC, IIAB, IID, IIC, IIF, IIG, IIIAB, IIIE, IIIF, IVA, IVB* | |
| Non-magmatic *IAB, IIE, IIICD* | IIEs related to H chondrites, IABs related to winonaites |

[1]A few anomalous eucrites may not be from 4 Vesta.
[2]15% of irons remain ungrouped.

Table 4. The average ages and estimates of the accretion ages of chondrites and various non-chondrites from Sugiura and Fujiya (2014) unless otherwise indicated. The ages are in millions of years after CV CAI formation.

| **Group** | **Chondrules** | **Time after CV CAI** | **Other** |
|---|---|---|---|
| **Chondrites:** | | | |
| E | | 1.8±0.1 | |
| O | 2.0+0.5,-0.3[a], 1.7±0.7[b] | 2.1±0.1, ≥2-3[c] | |
| R | | 2.1±0.1 | |
| CK | | 2.6±0.2 | |
| CO | 2.0+0.3,-0.2[d] | 2.7±0.2 | |
| CV | | 3.0±0.2 | |
| CI, CM, CR, TL | | 3.5±0.5 | |
| **Non-chondrites:** | | | |
| Angrites | - | 0.5±0.4 | ≤1.5e |
| HEDs | - | 0.8±0.3 | ≤2.5±1f, ≤0.6+0.5,-0.4g, <1h |
| Magmatic irons | - | 0.9±0.3 | 1.0±0.6, 1.3±0.5, 1.5±0.5i |
| Stony irons | - | 0.9±0.3 | |
| Ureilites | - | 1.0±0.3 | |
| Acap.+Lodran. | - | 1.3±0.3 | |
| Aubrites | - | 1.5±0.1 | |
| NWA 011 | - | 1.5±0.1 | |
| Tafassasset | - | 1.9±0.2 | |

[a] – see text for details. [b] - Kleine et al., (2008). [c] - Harrison and Grimm (2010); Henke et al., (2013); Kleine et al., (2008); Trieloff et al., (2003). [d] - see text for details. [e] – Kleine et al., (2012). [f] - Trinquier et al., (2008). [g] - Schiller et al., (2011). [h] - Neumann et al., (2014). [i] - Kruijer et al., (2013).

Appendix A: Major outstanding asteroid compositional questions

[See Excel File]

Acknowledgements:
We would thank the following authors for fruitful discussions: B. Carry, R. Brunetto, T. Burbine, P. Vernazza, D. Polishook, M. Elvis, C. Lisse, V. Reddy. We thank N. Moskovitz and G. Libourel for helpful reviews that improved the manuscript. This material is based upon work supported by the National Aeronautics and Space Administration under Grant No. NNX12AL26G issued through the Planetary Astronomy Program and through the Hubble Fellowship grant HST-HF-51319.01-A, awarded by the Space Telescope Science Institute, which is operated by the Association of Universities for Research in Astronomy, Inc., for NASA, under contract NAS 5-26555. This material is based upon work supported by the National Science Foundation under Grant No. 0907766 C. M. O'D. Alexander acknowledges partial support from NASA grant NNX14AJ54G and from NASA's Astrobiology Institute. K. J. W. was partially supported by the NASA Planetary Geology and Geophysics Program under grant NNX13AM82G.

Appendix

| Topic | Question | Possible Solutions | What Next? | Related AIV Chapters |
|---|---|---|---|---|
| | | | | |
| A. Formation & Physical Evolution of Asteroids | 1. Where, when, and how did they form? | *Where*: Links between asteroids & meteorites help constrain the conditions and location of formation. Dynamical models link where the asteroids are located now to where they originally formed. *When*: Recent advances dating mateorites have constrained formation ages. For example, igneous meteorites have crystallization ages older than chondritic ones. So bodies that melted, melted early. Bodies that did not melt formed later. *How*: A second leading theory on formation, through streaming instability, has emerged since Asteroids III. | Isotopic evidence points to carbonaceous chondrites being different from all others. But they are not all more water-rich, and their water H isotopes are not comet/Enceladus-like. Ages of chondrules and secondary minerals can be used to estimate chondrite formation ages. The almost ubiquitous presence of chondrules suggests that they may have played a role in planetesimal formation. | Section 4, this chapter, Johansen |
| | 2. How have they evolved physically? | Asteroid compositional changes are caused by early heating, acqueous alteration, and the external environment (space weathering). Physical changes are caused by impacts, radiation pressure (YORP spinup) and close approaches to planets (large tidal force). | Continued work on collisional and thermal modeling, studies of families, spacecraft measurements including surface (crater) observations. In-situ measurements and sample return of non-OC-like bodies. Density measurements to understand interiors. | Brunetto, Krot, Scheinberg, Scott, Wilson, Bottke, Asphaug, Jutzi |
| | 3. How have they evolved dynamically? | Planetary migration likely plays a leading role in early Solar System history. Planetary scattering, collisions, orbital drift, and orbital changes due to resonances dominate later history. | Identify the current spread in semi-major axis of each compositional group and use it to constrain the different migration scenarios. Search for and characterize interlopers for composition. Orbital and mass constraints also continue to inform dynamical history. | Section 3, this chapter. Morbidelli, Vokrouhlicky, Bottke |
| | | | | |
| B. Asteroid & Meteorite Compositions | 1. How many original parent bodies are represented in the asteroid belt? | Based on the meteorite record and the assumption that each meteorite group comes from a single parent body there are at least 100-150 distinct parent bodies. | Meteor observation networks will provide insight on source bodies for meteorites. Continued work linking large main belt asteroids to meteorites. | Burbine et al, AIII |
| | 2. Can a single meteorite group be represented by more than one parent body? | 1) The simplest assumption is that each meteorite group comes from one parent body.<br>2) More than one object could have formed at a similar time and a similar distance from the Sun, in which case they might look very similar. For instance, there could be more original (>50-100 km) S-type asteroids than the 3 needed to explain the three ordinary chondrite groups. | Further research on meteorite composition will yield better understanding of parent bodies. For example, high-precision isotopic measurements may reveal if the OCs separate and demand more than three bodies. Sample return will provide insight as will a review of how many primary parent bodies (those > 50-100 km) there are in each spectral class. | Burbine et al, AIII |
| | 3. How well do meteorites sample the asteroids? | Comparison of meteorites with micrometeorites and breccias suggests meteorites may be fairly representative samples of the major types of asteroids. We still find new types of meteorite (and presumably parent bodies), but differences tend to be subtle. However, the inherent "top-heavy" asteroid size distribution means that rare, large collisions stochastically dominate the ejected fragments by mass, resulting in an inherent possibility of large devations from direct representation of asteroid types by meteorite collections. If there is more diversity in the spectra of main belt asteroids than meteorites, could much of this diverity be due to regolith processes (grain size and density sorting) and space weathering? Also, there are biases associated with delivery efficiency from different resonances and the robustness of the samples as they past through Earth's atmosphere. | Characterize asteroids at sizes relevant to meteorite falls (~5-50m) to compare with the meteorite collection. IDPs, micrometeorites and clasts in meteorite regoliths provide alternative samplings of the asteroid belt - how similar are they to the meteorites? | Borovicka, Jenniskens, Binzel |
| | 4. How robust are our asteroid-meteorite links? | A few connections are robust: OCs make up part of the S class. HEDs are linked with V-types and the largest group of isotopically linked HEDs are concluded to be from Vesta. Isotopically distinct HEDs, as well as the diversity of other achondrites, point to a wide diversity of differentiation processes that remain poorly understood. The CMs may be linked with Ch and Cgh asteroids. The weaker and fewer bands present in an asteroid spectrum the less confident we are of its composition. The C and X complexes could be extremely compositionally diverse, but observations are also affected by varying grain size, phase angle, regolith gardening, space weathering etc. Shock darkening, which also mutes absorption bands, can also disguise the compositional identity of asteroid surfaces. | Dynamical study of asteroid families has the potential for addressing this question. By determining the ages of families and comparing with meteorite shock ages, and by following the plausible dynamical routes from family to Earth, current best guesses of associations of some meteorite types with families might become more robust. The mid-IR may be the next frontier for ground-based observational studies. Meteorite studies of spectral effects not related to composition are needed. Asteroid sample return will provide valuable insight for featureless asteroids. Serendipitous observation and recovery of objects such as 2008 TC3 will also provide "free sample return." | Vernazza, Brunetto, Yoshikawa, Reddy, Burbine AIII Chapter |
| | 5. How well do NEOs represent the Main Belt & beyond? | We now understand that dynamical and weathering processes can be relatively fast, suggesting that NEO flux is just a current snapshot, influenced by stochastic events like more recent disruption events. Size might also matter, the speed at which an asteroid's orbit drifts due to the Yarkovsky effect increases with proximity to the sun and with decreasing diameter. Yarkovsky is more effective at small sizes (10 meters or smaller) which might dominate meteorite samples. Additionally, size-dependent delivery mechanisms (Yarkovsky) mean that different size ranges could be dominated by specific asteroid families. NEO lifetimes and the NEO delivery models have helped link NEOs to their main belt source regions. | Survey Main Belt asteroids at sizes similar to NEOs (~1km). Study dynamical and compositional links between NEOs and Main Belt families and specific regions. Dynamics need to be calibrated by observations. New understanding of differentiation processes is also relevant. | Binzel |
| | 6. What is the diversity of compositions within individual small asteroids? What processes mix wildly different meteorite types into a single tiny body (e.g. Almahatta Sitta, Kaidun)? When did the mixing occur? | Collisional, accretional, or both processes could potentially bring such diverse materials together. | Implementation of ATLAST-like telescopic surveys of asteroids/meteoroids on their final approach to Earth and increased video surveilance and recovery of fall samples to understand the prevalence of and compositions of these mixes. Physical measurements of the smallest asteroids (5-100 meter size). Sample return of small asteroids will also provide constraints. | Section 5.3 This Chapter, Borovicka, Bottke? |
| | 7. The Ordinary Chondrite Paradox: Why does the most common asteroid type S-type not match the most common meteorite type OC? | Space weathering is the primary reason for the spectral mismatch. Lab experiments plus ground- and space-based asteroid measurements made great progress. Hayabusa's sample return of Itokawa provided conclusive evidence. Other factors affecting spectral slope include grain properties and observational phase angle. | This question is solved. The follow up questions are: What S-type asteroids are not OCs? What meteorites do they supply? How does the space environment affect other asteroid types? | Sec 5.5 This Chapter Vernazza, Brunetto, Binzel, Yoshikawa |
| | 8. What are the interior compositions of asteroids? | Density measurements and asteroid families currently provide the most information about asteroid interiors. How compositionally homogenous or differentiated the medium-to-large asteroids are is largely unkown. | Density measurements particularly from multiple systems. Although the porosity of asteroid interiors must be better constrained as well. | Margot, Scheeres, Barucci |
| | | | | |
| | 1. What is the source of the compositional gradient in the | 1) It is a primordial remnant from the temperature and compositional gradient in the disk. | [illegible] | Section 2.2, 2.3, |
| | main belt? Why are the Hildas and Trojans compositionally homogeneous compared to the Main Belt? | 2) It is the result of a transplantation of one or more groups of asteroids that formed elsewhere. 3) Hildas and Trojans actually are more compositionally diverse than they appear, but they have significant low-albedo materials that renders diagnostic spectral features nearly invisible. | Progress on early solar system environment models and asteroid formation models. The best though impractical way to solve this is a mineralogical and isotopic assay of dozens of asteroids and comets. | 5.1 This Chapter, Morbidelli, Johansen, Emery |
| **C. Asteroid Compositional Distributions** | 2. How does the distribution of asteroids change as a function of size? What is the significance of that distribution? | Recent work has explored the change in relative abundance of asteroid types as a function of size. Many factors still need to be taken into account such as: 1) the size-frequency distribution of families 2) the difference in collisional lifetimes per asteroid class 3) Some compositions are masked at smaller sizes due to processes such as collisions and "shocking" | Additional study of the size distribution of families in the inner belt. Ground-based imaging and shape models plus mission visits to primitive bodies. Constrain how prevalent shocking is in the main belt. Larger samples of small asteroids in the MB (1-20km) will help determine the distribution at smaller sizes. | Section 5.2 This Chapter |
| | 3. The Missing Mantle Problem: Where is all the missing mantle material? Additional Questions: Why are V- and A-types scattered throughout the entire main belt? | 1) Asteroids differentiate differently than expected - maybe don't form large olivine-rich mantles and pyroxene-rich crusts.<br>2) The parents of these cores formed in the terrestrial planet region. They were destroyed and only the strongest metallic fragments were subsequently delivered to their current locations in the main belt.<br>3) "Battered to bits" - this theory is currently less favored. Collisional modeling, crater counts, and observational evidence do not support an aggressive regime of collisional destruction and battering.<br>4) Previous theories postulated olivine was hidden by weathering processes. Recent progress on on space weathering disproves this hypothesis. | Further study of dynamical solutions and differentiation modeling. Continued study of meteorites to understand differentiation. Continued search for differentiation in families, including metal within large families. | Section 5.4 This Chapter, Elkins-Tanton, Wilson, Scott |
| | 4. Where is the water in the asteroid belt? How much is there? Where did these water-rich asteroids form? | Current evidence: main belt comets, activated asteroids, water absorptions, Ceres outgassing, and possible exposures of ice on Ceres. | Discover additional active asteroids and explore asteroid-comet connections. Continue studies of extinct or dormant comets among NEOs. Visit and map surfaces such as by missions DAWN, OSIRIS-REx, and Hayabusa II. Also, density measurements, radar sounding by spacecraft, etc. might reveal ice buried beneath thick surficial lag deposits. | Jewitt, Rivkin, Krot, Binzel |
| | | | | |
| **D. Asteroids in their Greater Context** | 1. Asteroid belt's importance for Earth - do we have remnants of Earth's building blocks? Do we have remnants of Earth's water source? | Arguments exist for and against enstatites or angrites being primary components of Earth. Water from Earth's oceans are argued to have been delivered from asteroids. Late veneer. Isotopic compositions show that the known chondrites were not the major building blocks of Earth, but the CI+CM chondrite-like material may have been the sources of Earth's volatile materials as suggested by SMOW D/H ratio. | Continued studies of D/H ratios of meteorites, comets, and asteroids. Continued theoretical studies of planet formation and dynamical studies of planet and planetessimal migrations. | Section 6.1, this chapter |
| | 2. Asteroid belt's importance for rest of our SS? | Surface ages determined by impacts and cratering rates. Main Belt and Kuiper Belt orbital architecture and captured Trojans and satellites constrain giant planet migration. Study of solar corona is enabled by asteroids and comets evaporating during close approach to the Sun. | Continue with current progress on crater studies, dynamical studies including migration. | Beyond the scope of this book. |
| | 3. What role do asteroids play in other planetary systems? | Collisions in massive asteroid belts create debris disks around old stars<br>Asteroids are tracers of dynamics and physical properties in Dusty White Dwarf systems. | Identify and characterize extrasolar asteroid belts (evolved debris disks). Survey different planetary system architectures including asteroid belt distances and masses.<br>Identify and characterize dusty white dwarf systems, and link those systems to the expected properties of precursor planetary systems. | Section 6.2 this chapter. Mostly beyond the scope of this book. |
| | 4. What role do asteroids play in creating the conditions for habitability on Earth-size planets? | Meteorites are known to contain water and organics (including amino acids), which could be precursors for life. They represent the initial conditions and compositions for forming terrestrial planets. They could be responsible for water delivery and creating oceans. They could also be responsible for destroying life through impacts. | Understand the link or the gradient between comets and wet asteroids. Measure the D/H ratio of a wider array of water-rich small bodies. Dynamical studies of delivery. Understand consequences of delivery of too much water. Studies of impact and water retention on Earth. Studies of ocean-forming mechanisms unrelated to small body delivery. | Beyond the scope of this book. |
| | 5. What are asteroids' importance as hazards and resources to Earth? | Asteroids as a hazard have motivated increased attention to discovery surveys. Asteroids contain rare-Earth elements including metals of value. Asteroids contain water or water components useful for space travel. | Discover 90% of PHAs down to 140meters. Understand the size of the small NEA population (<140 meters). Evaluate and prepare for deflection strategies or evacuation plans. Continue to explore asteroids as resources. | Harris, Jedicke, Farnocchia |
| | | | | |